# Aquila

**QuEra's 256-qubit neutral-atom quantum computer**

Version 1.0, June 20, 2023

**Algorithms & Applications team:**

**Jonathan Wurtz**, Fangli Liu, Phillip Weinberg, John Long, Sheng-Tao Wang

**Aquila team:**

**Alexei Bylinskii**, Boris Braverman, Sergio H. Cantu, Florian Huber, Jesse Amato-Grill, Alexander Lukin

Nathan Gemelke, CTO
Alexander Keesling, CEO

QuEra Computing Inc.
White Paper

Target Audience:
Prospective Users
Quantum Educators

# Selected "best practices."

In this whitepaper we include many suggestions on how to use Aquila to the best of its capabilities. A few of our favorite recommendations are collected here.

### On post-selecting on fully filled arrays

When analyzing data from Aquila, it is important to "post-select" on a correctly filled atom array, else you may occasionally get unexpected results. For implementation details, see the example notebooks [here](#).

- Alexei Bylinskii

### On designing smooth waveforms

The optical control elements have a large but finite bandwidth, which may cause rapidly varying waveforms to lead to unexpected behavior. Consider designing smooth protocols wherever possible.

- Boris Braverman

### On "parallelizing" to reduce shot counts

When doing few-atom dynamics, you may reduce the number of shots by parallelizing the same configuration multiple times across the array. These examples typically use a 4×4 array of atoms spaced 25μm apart.

- Phillip Weinberg

### On maximizing Rabi frequency

When implementing dynamics, it is important to keep the protocol as short as possible to minimize decoherence effects. If possible, choose the maximum possible Rabi drive $\Omega$ to minimize the time given fixed total pulse area $\Omega t$.

- Nathan Gemelke

### On implementing phase jumps

When executing a large jump in phase, it is best to not have the Rabi drive active. This is due to the particulars of the AOMs that drive the phase and amplitude of the Rabi drive.

- Alexander Keesling

### On the robustness of adiabatic protocols

Adiabatic protocols are a flexible and robust mode for analog computation, due to a relative insensitivity to phase, amplitude, and position noise. Consider them when designing analog algorithms!

- Jonathan Wurtz

### On positioning atoms away from the blockade radius

When designing protocols and algorithms, it is best to set atoms either deep within or far from the blockade radius; otherwise, the evolution may be sensitive to thermal position fluctuations.

- Shengtao Wang

### On choosing the number of shots

There is a trade-off between shot noise and speed/cost. ~100 measurements are a good middle ground, with up to 1000 shots needed for high-resolution phase diagrams and low-probability outcomes, and as few as 25 shots for parallelized few-atom arrays.

- Fangli Liu

# Table of Contents

# 1. Introduction

The neutral-atom quantum computer "Aquila" is QuEra's latest device available through the Braket cloud service on Amazon Web Services (AWS). Aquila is a "field-programmable qubit array" (FPQA) operated as an analog Hamiltonian simulator on a user-configurable architecture, executing programmable coherent quantum dynamics on up to 256 neutral-atom qubits. This whitepaper serves as an overview of Aquila and its capabilities: how it works under the hood, key performance benchmarks, and examples that demonstrate some quintessential applications. Complementary to the whitepaper are a series of Jupyter notebooks which implement each example using the Braket SDK, and are available via [QuEra's GitHub page](#).

This whitepaper is intended for readers who are interested in learning more about neutral-atom quantum computing, as a guide for those who are ready to start using Aquila, and as a reference point for its performance as an analog quantum computer. We assume basic knowledge of quantum mechanics (Hamiltonians, Hilbert spaces, qubits, etc.), but no further advanced concepts are needed.

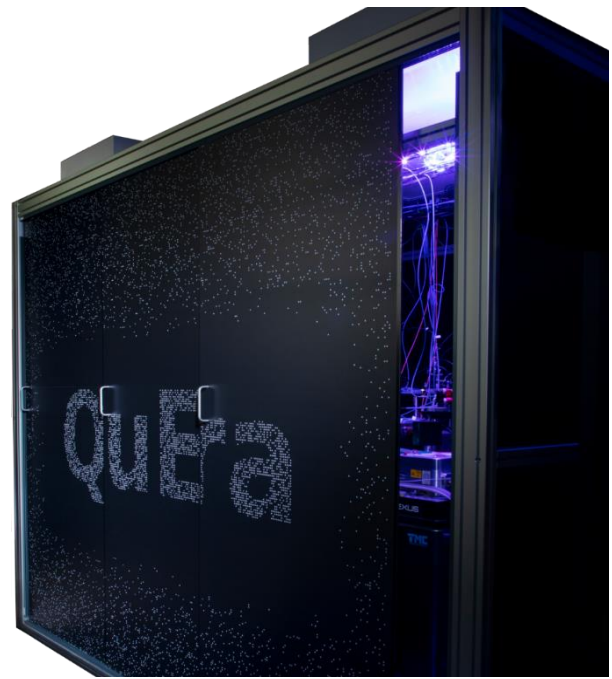

The exterior of Aquila, a "field-programmable qubit array" (FPQA) operated as an analog Hamiltonian simulator that implements quantum computations with up to 256 neutral-atom qubits.

A core philosophy of QuEra is transparency and managing expectations of our hardware and solutions. Quantum computing has the potential to change the world by tackling intractable computational problems by leveraging quantum physics and its exponentially large computational space. Unfortunately, today's noisy intermediate scale quantum (NISQ) era devices are limited in generality and problem size. This whitepaper will showcase Aquila through data, to highlight both its strengths and limitations. Even though Aquila has noise and decoherence limitations, there is clear evidence of quantum effects that underlie its operation and its performance. While imperfect, we believe that the noisy quantum computation can be leveraged for near-term applications and algorithms that have a real-world impact. At the same time, we understand that Aquila is only the first step on our long journey towards large-scale fault-tolerant quantum computing.

The rest of this whitepaper is separated into two parts. Part one goes over the basics of neutral-atom quantum computing. Part two introduces five examples of increasing complexity from single-qubit dynamics to combinatorial optimization. Each section also has an [associated Jupyter notebook](#) which can be run through Amazon Braket to reproduce these results. All data was acquired on Aquila on June 9, 2023. The data presented in this whitepaper total 96,250 measurements over 1,884 tasks.

If you are interested in learning more about Aquila or other quantum computers being developed by QuEra, feel free to reach out to us. If you wish to get started running tasks and algorithms on Aquila, a good start is to look over the [example Jupyter notebooks](#). We also provide a private "premium access" mode with extended machine capabilities and dedicated support. If you are interested in these features, feel free to reach out to our sales team at [sales@quera.com](mailto:sales@quera.com).

## 1.1. Background and literature

While the teams at QuEra are working hard to make rapid advancements in this technology, our work stands on the shoulders of giants and is enabled by the incredible progress of many academic, government, and industry research groups over the last decade. Beginning with early work [Jaksch2000] proposing neutral atoms as a quantum computing platform and following a series of proof-of-concept experiments with Rydberg atoms as qubits [Isenhower2010, Wilk2010], the field has recently advanced rapidly after seminal papers demonstrating deterministic loading of large arrays of neutral-atom qubits using optical tweezers [Endres2016] and nonequilibrium dynamics of a 1D chain of 51 atoms [Bernien2017], highlighting the platform's capabilities as an analog quantum simulator. Recently, there have been a further series of advancements including high-fidelity two qubit gates [Levine2019, Evered2023], preparation of large entangled states [Omran2019], demonstrations of quantum phases in 2D [Ebadi2020], realizations of topological quantum phases [Léséleuc2019] including the first observation of a quantum spin liquid [Semeghini2021], application to classical optimization problems [Ebadi2022], quantum circuits with arbitrary connectivity using coherent atom shuttling [Bluvstein2022], and many more.

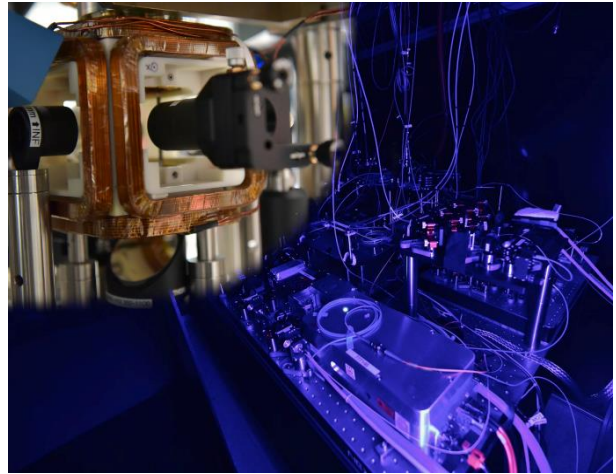

Figure 1.1 The interior of Aquila. An array of optical elements, lasers, and cameras (bottom right) focus on a vacuum cell (top left insert) filled with a dilute gas of Rubidium atoms. In an area less than three human hairs wide, the laser fields are carefully controlled to manipulate the state of up to 256 qubits to execute quantum computations.

These advancements highlight the promise of neutral atoms as a powerful quantum computing platform and several neutral-atom quantum computing companies have been spun off from the success of academic research labs. These include companies such as Pasqal, ColdQuanta, Atom Computing, Planqc, M Squared, and of course QuEra Computing, with more being added every year.

For more details on neutral-atom quantum computing, we recommend the reviews [Saffman2010] and [Morgado2021].

## 1.2. The key ingredients of neutral-atom quantum computing

Aquila is a room-temperature quantum device with neutral Rb-87 atoms held and cooled to microkelvin temperatures by laser beams inside a vacuum cell. The individual atoms are intrinsically quantum, encoding qubits in their electronic states and manipulated by precisely controlled laser pulses. These states are used to process quantum information and can be detected by state-dependent fluorescence. The core of Aquila is a 2-cm scale glass vacuum cell, a microscope objective, and low-noise camera, surrounded by mostly off-the-shelf optical components delivering the laser beams to the atoms, and data-center style racks filled with control electronics for laser beams and data acquisition systems. The microscope objective focuses light into an area less than $200 \mu m$ wide --- about the width of three human hairs --- inside the glass cell, where the atoms themselves are arranged in a 2D pattern.

For Aquila to do quantum computation, it must leverage four key ingredients: 1) individual Rb-87 atoms as physical qubits, 2) reconfigurable optical tweezer arrays to enable arbitrary arrangements of up to 256 atoms, 3) ultra-stable lasers that drive quantum dynamics in the atoms' electronic orbitals, and 4) Rydberg states that enable the atoms to interact with each other.

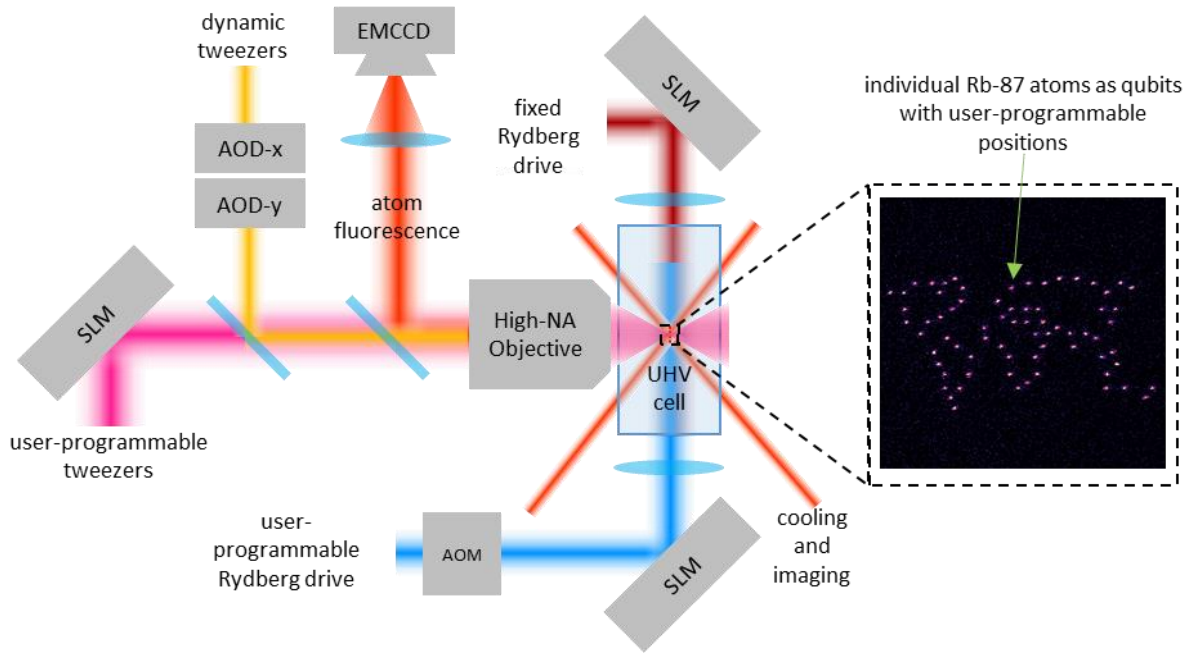

Figure 1.2. Functional block diagram of Aquila. There are six wavelengths of laser light that focus on the computational area in the vacuum cell. One laser (pink) is controlled by a spatial light modulator (SLM) to arbitrarily position up to 256 atom traps. Another laser (yellow) uses a crossed set of acousto-optic deflectors (AOD) to dynamically move atoms in traps and deterministically sort the array. A set of lasers (red) is used to cool the atoms to $\mu$K temperatures, and two counter-propagating lasers (deep red and blue) implement a two-photon drive between ground and Rydberg state. A final laser and camera (orange) is used to image the position of atoms in each trap using fluorescence. An example image of individual atoms leveraging arbitrary positioning is shown to the right.

# Key ingredient: Rubidium atoms

In Aquila, each qubit is physically a single Rb-87 atom, which needs to be isolated, cooled, moved, initialized, manipulated, and measured. Electronic orbitals of the valence electron of the atom encode the quantum information and other intermediate orbitals are used to manipulate the state. A simplified electronic structure diagram is shown in figure 1.3.

There are two sets of states that are useful to represent a qubit on the Rb-87 atom: the **ground-Rydberg qubit**, which is relatively short-lived but enables strong interactions between qubits which are responsible for entanglement, and the **hyperfine qubit**, which gives no interactions but is a long-lived memory qubit. Having both qubit types hosted by a single atom, and the ability to switch between them using laser pulses, is a tremendous advantage of the neutral-atom platform. While Aquila currently operates with the ground-Rydberg qubit only as an analog quantum computer, the hyperfine qubit will be a key ingredient for a later digital gate-based or **dual-mode** (analog/digital) operation.

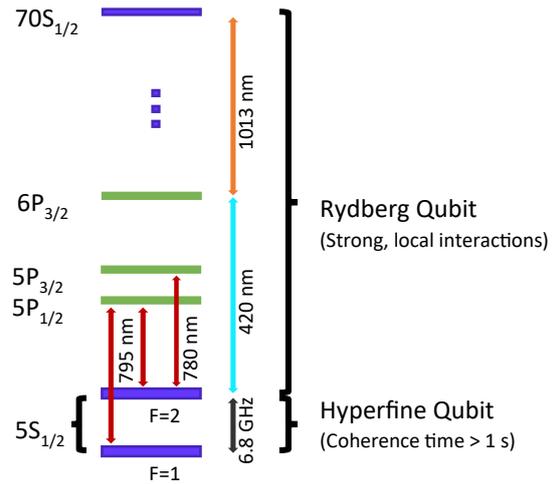

Figure 1.3. Rb-87 valence electron states utilized in Aquila to manipulate the atom as a qubit and as a physical host to the qubit. Arrows are the various optical fields used to drive transitions. Purple lines are the states that represent the qubit, while green lines represent other states used in the various manipulations.

The hyperfine qubit is represented by two hyperfine ground states $|0\rangle = |g\rangle = |5S_{1/2}, F = 1\rangle$ and $|1\rangle = |g'\rangle = |5S_{1/2}, F = 2\rangle$ separated in energy by a transition frequency $\approx 6.8$ GHz. These qubits have the advantage of extreme stability and long coherence ($\sim 1$ second) due to weak interactions with the environment and other qubits. To execute entangling gates, the atoms are temporarily excited to the Rydberg state in a way that is robust to errors. Results from 2019 find 2-qubit entangling gates with 97.4% fidelity [Levine2019], while recent results demonstrate a high fidelity of over 99.5% [Evered2023].

For the Rydberg qubit, the $|0\rangle \equiv |g\rangle = |5S_{1/2}\rangle$ is represented by a ground state, while the $|1\rangle \equiv |r\rangle = |70S_{1/2}\rangle$ is represented by a highly excited S orbital, called the Rydberg state. This Rydberg state has many useful properties that can be used to generate entanglement, as will be described later.

Rydberg basis measurements are implemented through re-trapping the atoms. During quantum evolution, the optical traps are turned off so that they do not affect the quantum dynamics. After evolution, the optical traps are turned back on, which collapses the wavefunction to a particular logical basis state. The ground state is re-trapped by the lasers, while the Rydberg state is anti-trapped by the lasers, quickly pushing the atoms out of the array. The state is then measured through fluorescence imaging through the absence or presence of an atom: No atom (0) means Rydberg state, while the presence of atom (1) means ground state. This measurement process has some errors, as the traps may spuriously recapture a decayed Rydberg state or fail to recapture a ground state. This effect is especially evident in example 3, which measures the many-body fidelity of a 1D chain.

Measurements are "destructive", in that atoms are lost every time they are measured to be in the Rydberg state, which means that every experimental cycle rebuilds the atom array from scratch. While this process is relatively slow (<10 Hz), it highlights the arbitrary reconfigurability of the platform, as a completely different geometry can be in principle chosen from shot to shot, effectively building a different quantum processor at each shot.

# Key ingredient: FPQA with arbitrary geometry and optical tweezers

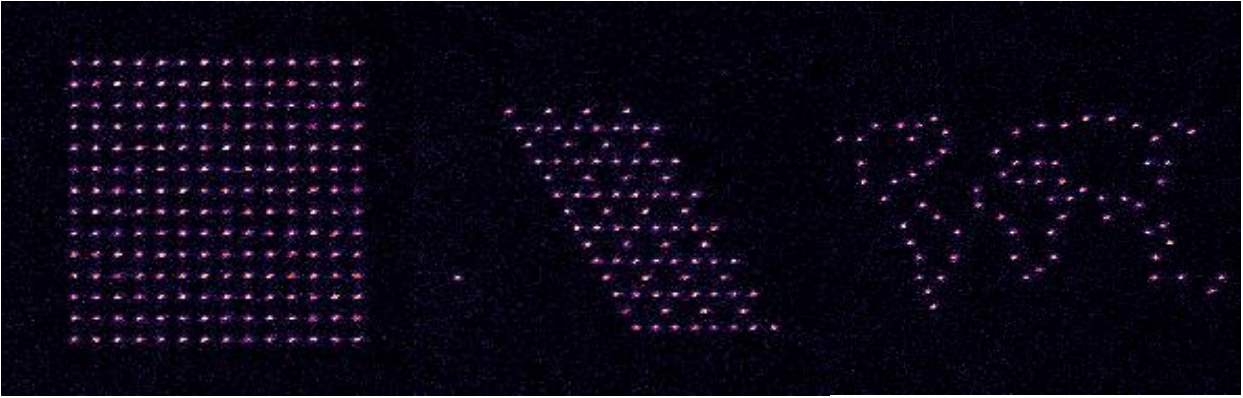

Figure 1.4. Examples of arbitrarily positioned atom arrangements enabled by reconfigurable tweezers: Left: a regular array of qubits as a quantum register in a gate-based architecture, Middle: qubits arranged in a Kagome lattice to encode a quantum simulation problem, Right: qubits arranged in the shape of the world coastlines to encode a geographical optimization problem.

Aquila is a field-programmable qubit array. An integral part of the FPQA capability is to trap and rearrange individual atoms using focused laser beams known as **optical tweezers** by way of the optical dipole force. This process, also known as **optical trapping**, uses a laser tuned close to resonance with some intermediate state (in the case of Aquila, a 780nm laser tuned to the $6P_{3/2}$ transition of Rubidium) to induce a dipole moment in the atom. In turn, this induces radiation pressure towards the area of high intensity. By tightly focusing the laser on a single spot, the neutral atom becomes trapped. A second set of lasers optically cool the atoms by converting kinetic energy into photonic energy and initialize every atom into the ground state.

Aquila uses optical trapping in two modes. The first mode creates a quasi-static array of hundreds of individual traps, using a device called a **spatial light modulator** (SLM). An SLM, which uses technology similar to a typical presentation projector, employs an array of liquid crystals to set the phase of the laser wavefront as a function of position in the Fourier plane. By carefully engineering this phase mask in a process known as **holography**, the laser field forms hundreds of tight spots on the focusing plane, which serve as the locations of traps for individual atoms. While the response time of the SLM is too long for the microsecond timescales of quantum evolution, the phase mask can be changed in between measurements to set the positions of traps, and thus atoms, into any **arbitrary**

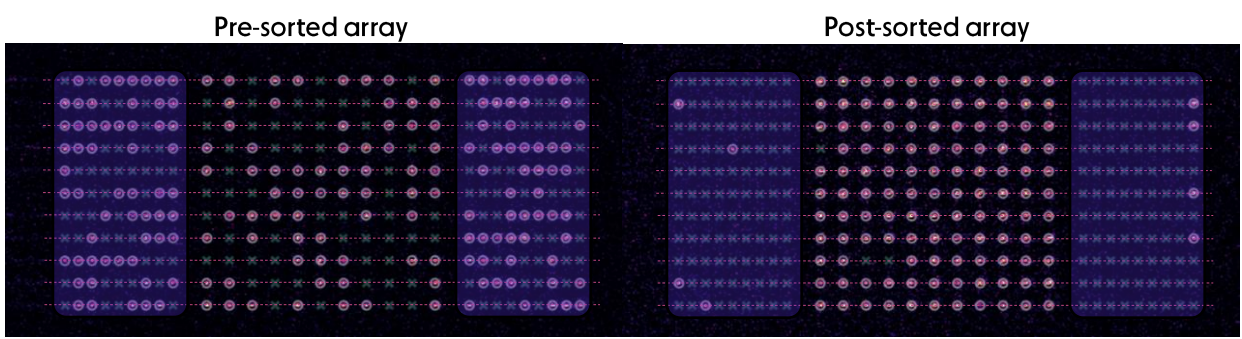

Figure 1.5. The deterministic loading process of an 11 × 11 square lattice. On the left is an image of the pre-sorted, stochastically loaded array; empty sites are indicated with an ×, and filled sites with ∘. Using a laser tweezer, atoms are moved from the reservoir regions (purple) on each side to the user region in the center to create a deterministically loaded array on the right image. Note that the process is sometimes imperfect due to incomplete filling of rows and atom loss, as is clear from the three empty sites of the post-sorted array. Observe the clear requirement that each atom in the array be a part of a row (red dashed line). Note that only the user region in the middle is visible to users on Amazon Braket.

**geometry**. Some examples of arbitrary placement are shown in Fig. 1.4, which image the fluorescence of atoms held in these traps. While in principle the positioning of atoms is arbitrary, in practice it is subject to physical constraints. On Aquila, no two sites can be placed within $4\mu m$ of each other due to the resolution of the SLM, and the area of where traps can be placed is constrained within a $75\mu m \times 76\mu m$ square due to the size of the focusing optics. These constraints must be an important consideration when designing atom placements and may be relaxed with "premium access" mode.

The second mode of optical trapping uses a set of acousto-optical deflectors (AOD) to dynamically move atoms. An AOD uses an acoustic wave traveling across a crystal to create a diffraction grating, which deflects light by an amount controlled by the frequency of the acoustic wave. By using two crossed AODs, one can control the location of several trapping spots in the atom array and move them on microsecond timescales. This enables moving atoms between traps, which is crucial to deterministically loading Aquila's atom arrays, as well as future dynamical reconfigurability and arbitrary-connectivity digital architectures based on mid-circuit atom shuttling [Bluvstein2022]. The preparation of a deterministically loaded atom array in Aquila happens in several steps and is shown in Fig. 1.5. First, atoms are captured from a dilute vapor at room temperature. Then, the static SLM traps are turned on and each is filled with a single atom with a ∼60% probability. About 600 total traps are required to load a maximum of 256 atoms in a square lattice.

> **Best Practices tip:**
> When analyzing data from Aquila, it is important to "post-select" on a correctly filled atom array, else you may occasionally get unexpected results. For implementation details, see the example notebooks here.

These extra traps are added in a "reservoir area" on each side of the array. After taking an image to determine the locations of each atom, the dynamical AOD laser tweezer decides and executes a series of moves that load atoms from the randomly filled sites to the user-specified sites. Because of this dynamic and real-time process, the atoms are sorted with very high probabilities over 99%. However, as part of the analysis, one typically needs to post-select on a fully filled array; for more details, see the example Jupyter notebooks.

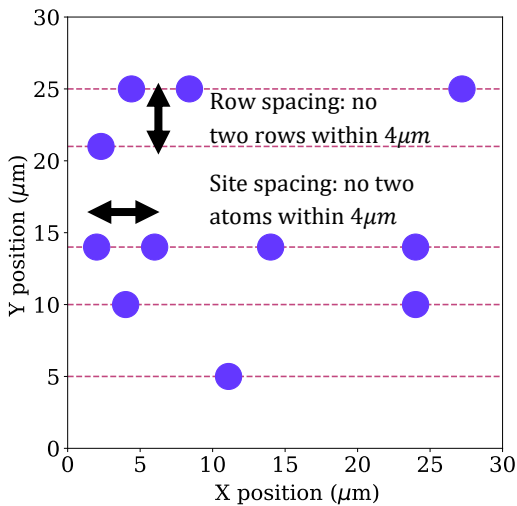

Figure 1.6. Atom geometry constraints. Sites (purple) must be placed in rows (red dashed). Rows must be at least $4\mu m$ apart, and each site must be spaced at least $4\mu m$ apart.

> **Best Practices tip:**
> The geometry requires that every site be set in a row, which means that the Y coordinate must be either equal or at least $4\mu m$ away from all other sites. Arbitrary placements may need to be modified to fit this constraint.

To accelerate the sorting process, atom locations are required to be laid out in discrete rows, as can be seen in the horizontal red dashed lines in Fig. 1.5 and Fig. 1.6. Atoms within each row are then simultaneously rearranged into the desired final structure. This imposes a strong constraint on valid positions of each atom in the array: atoms must be set in rows that must be at least $4\mu m$ apart, which means that truly arbitrary positionings must usually be relaxed into nearby rows. Furthermore, sites must be no closer than $4\mu m$ from any other site due to optical resolution, as seen in Fig. 1.6. These constraints are less impactful on regular lattices, such as the Kagome lattice, but may be an important consideration when designing arbitrary atom placements. The row restriction was a design decision to speed the sorting process, which is in no way fundamental and may be relaxed in "premium access" mode.

# Key ingredient: Photonics, lasers, and analog control

Transitions between different electronic states are implemented by absorption and emission of photons. When an atom is in the presence of light with the same energy as a particular energy transition, it (quantumly) absorbs or emits a photon to change between a ground and excited state. For this process to be coherent, the photons' electromagnetic field must be frequency stabilized to within ~ 10 kHz, less than 1 part in $10^{11}$, and is generated by ultra-stable lasers frequency locked to a cavity. This is a challenge of neutral-atom quantum systems: to have a high-fidelity quantum computer, these lasers must be simultaneously ultra-stable and deliver high power. A recent enabling technology has been the development of such lasers, which are highly stable, have the suitable wavelength, and have powers of order 100s of watts.

> **Derivation of the atomic Hamiltonian in the presence of a laser**
>
> *Consider an atom described by a Hamiltonian with two electronic energy levels $|a\rangle$ and $|b\rangle$. There are two terms in the Hamiltonian: the first is the coupling term from the dipole moment d between states caused by the time-dependent electric field $E(t) = E_0 \cos(\omega_0 t + \phi)$ of the laser operating at angular frequency $\omega_0 = 2\pi f = 2\pi c/\lambda$. The product of the electric field and the dipole moment is typically denoted by $-E_0 d = \Omega$, the **Rabi frequency**. The second term is the difference in the energies $\omega_a$ and $\omega_b$ of the states, typically measured as the ionization energy of the electron (in eV or MHz)*
>
> $$H = \Omega(\,|a\rangle\langle b| + |b\rangle\langle a|\,)\cos(\omega_0 t + \phi) + (\omega_b - \omega_a)|b\rangle\langle b|.$$
>
> *This Hamiltonian is highly time dependent and can be simplified, assuming $|\omega_b - \omega_a| \approx \omega_0 \gg \Omega$. First, we go to the **rotating frame** $|b\rangle \mapsto e^{i\omega_0 t}|b\rangle$, which adds an inertial diagonal term to the Hamiltonian*
>
> $$H = \Omega(\,e^{-i\omega_0 t}|a\rangle\langle b| + e^{i\omega_0 t}|b\rangle\langle a|\,)\cos(\omega_0 t + \phi) + (\omega_b - \omega_a - \omega_0)|b\rangle\langle b|.$$
>
> *Next, we use the identity $2\cos(\omega_0 t + \phi) = e^{i\omega_0 t + i\phi} + e^{-i\omega_0 t - i\phi}$ to expand*
>
> $$H = \frac{\Omega}{2}\left(e^{i\phi} + e^{-i2\omega_0 t - i\phi}\right)|a\rangle\langle b| + h.c. + (\omega_b - \omega_a - \omega_0)|b\rangle\langle b|.$$
>
> *Observe that the exponentials are sums or differences of different frequencies. Choose $\omega_a - \omega_b + \omega_0 = 0$ to set one to zero frequency, which means the second exponential is $e^{-2i\omega_0 t}$. By the **rotating wave approximation,** this extremely high frequency component can be neglected. Finally, the Hamiltonian becomes*
>
> $$H = \frac{\Omega}{2} e^{i\phi}|a\rangle\langle b| + \frac{\Omega}{2} e^{-i\phi}|b\rangle\langle a| + (\omega_b - \omega_a - \omega_0)|b\rangle\langle b|.$$
>
> *The **Rabi frequency** $\Omega$ is the characteristic frequency at which the atom is driven between states $|a\rangle$ and $|b\rangle$. The value $\omega_b - \omega_a - \omega_0 = -\Delta$, which represents how off-resonant the laser is from the atomic energy transition, is called the **detuning $\Delta$**. The value $\phi$, which is set by the time offset of the laser drive, is called the **phase $\phi$** and can always be set to zero at the beginning of the quantum evolution by $U(1)$ symmetry. While this derivation was done in a time-independent setting for only two states of one atom, it naturally extends to more states, atoms, and time-dependent parameters $\Omega, \Delta, \phi$ under reasonable assumptions on the laser modulation bandwidth.*

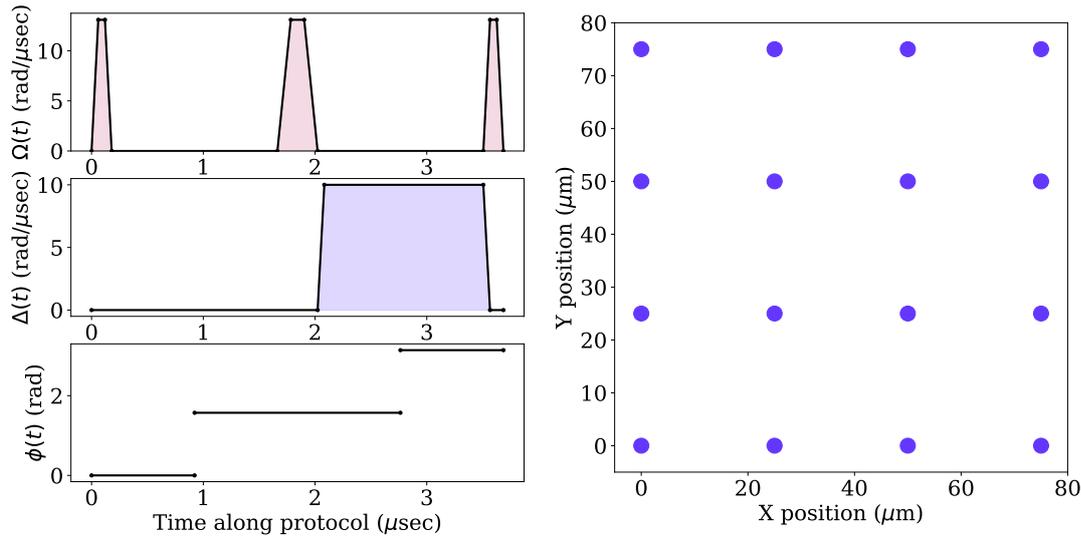

Figure 1.7. An example analog program on Aquila, defined by four quantities. First, the Rabi drive $\Omega(t)$, which sets the amplitude of the laser and thus the transition rate between ground and Rydberg state. The second is the detuning $\Delta(t)$, which determines how close the laser drive is to the atomic transition. The third is the phase $\phi(t)$, which analogously sets if the drive is in the X or Y direction. The fourth is the position $\{\vec{x}\}$ of each atom in the 2D array. This figure plots a program from example 1 implementing a spin-echo protocol.

Aquila's Rubidium-87 atoms use several wavelengths of lasers to implement quantum dynamics, as shown in Fig. 1.3. The energy difference between ground and Rydberg state is in the ultraviolet range, $\lambda \sim 300\ nm$, which is difficult to directly generate with a laser. Instead, the drive between ground and Rydberg state is implemented in a two-photon transition — $420\ nm$ and $1013\ nm$ — mediated by an off-resonant intermediate state $|6P_{3/2}\rangle$. In conjunction, these two lasers drive the transition between ground and Rydberg states at a characteristic rate given by the Rabi frequency $\Omega$, which is a function of the lasers' amplitude. The frequency of the lasers can be precisely tuned so that the energy of the photons almost exactly matches that of the transition between ground and Rydberg states. The offset from resonance is called the detuning $\Delta$. Execution of quantum computations and dynamics are implemented by precise control of optical fields, by choosing the amplitude, detuning, and phase of the laser as a function of time using optical components called acousto-optical modulators (AOM). These components use sound waves propagating in crystals to create a diffraction grating, with a response time and profile dependent on the speed of sound across the crystal (typically on the order of nanoseconds). Light propagates through the crystal in a way engineered to set a particular intensity, phase, and detuning.

**Best Practices tip:**
The optical control elements have a large but finite bandwidth, which may cause rapidly varying functions to lead to unexpected behavior. Consider designing smooth protocols wherever possible.

**Due to the arbitrary control of amplitude, phase, and detuning, Aquila is designed to work in a different way than the prototypical quantum computer, called the "Analog mode".** Instead of specifying a sequence of gates to execute quantum programs, a user instead defines a time series of the Rabi drive rate $\Omega(t)$, detuning $\Delta(t)$, phase $\phi(t)$, and position of each atom $\{\vec{x}\}$. An example program on Aquila is shown in Fig. 1.7. Unlike the prototypical "digital mode" computation, the interaction between each qubit is always on, which means that entanglement and correlations can more quickly build up throughout the system. However, this also means that computation is not necessarily universal, and is specialized for certain tasks such as optimization, simulation, and machine learning. **Analog mode is a key distinguishing feature of Aquila** over other contemporary quantum computing systems.

# Key ingredient: Rydberg states and the Rydberg blockade

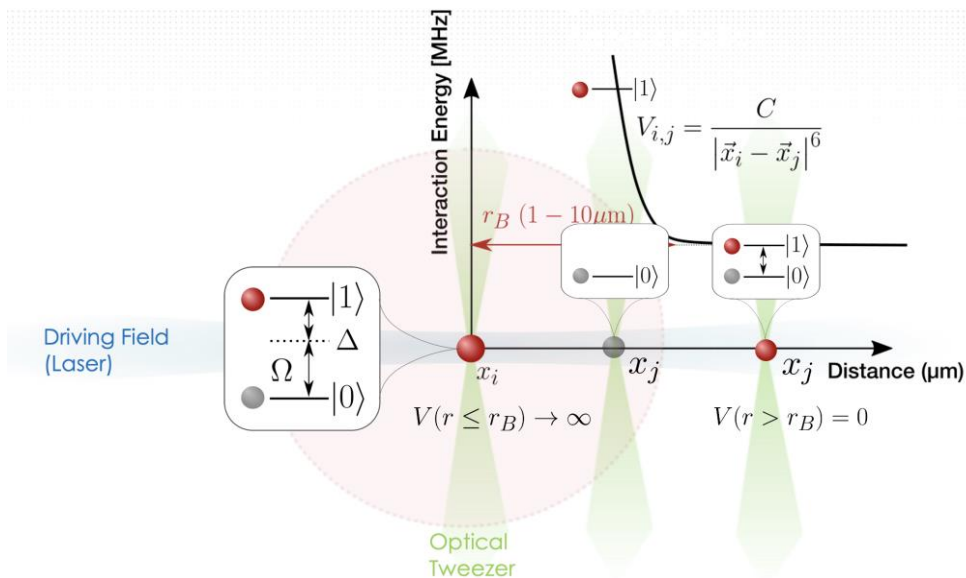

Figure 1.8. The Rydberg blockade mechanism. Two atoms are at some distance away from each other, where atom $i$ is in the Rydberg state. Outside of the blockade radius (red), atom $j$ can freely be driven to the Rydberg state. Inside the blockade radius, the Rydberg state is significantly detuned from the driving laser due to the strong interactions between nearby Rydberg-state atoms, preventing the atom $j$ from going into the excited state. This behavior is independent of the specific position of the atoms, and so entanglement can be generated robustly not just through the specific values of the *interactions*, but in the *structure of the Hilbert space*.

The final key technological ingredient of neutral-atom quantum computing is the **Rydberg state**. These states are highly excited electronic orbitals, which "puff" the valence electron to a large volume around the atom. The Rydberg state has a strong energy shift conditional on the state of adjacent atoms, which enables entangling dynamics. If two atoms are close together (on a length scale of order $\mu m$), they interact through a state-dependent energy shift. If only zero or one atom is in the Rydberg state, there is no energy shift. If both atoms are in the Rydberg state, there is an energy shift from the **Van der Waals interaction**, which depends on the sixth power of the distance

$$V_{ij} = \frac{5{,}420{,}503 \frac{\mu m^6 rad}{\mu s}}{|\vec{x}_i - \vec{x}_j|^6}$$

where the numerator is the $C_6$ constant for interaction between two $|70S\rangle$ states. As a characteristic scale, for $R = 8\ \mu m$, $V_{ij} \approx 20$ rad/$\mu$s. Due to the large exponent $R^{-6}$, there is a large falloff in the interaction as a function of distance. If two atoms are close together (say, 4 $\mu$m), the energy of the doubly excited state (e.g., 1,320 rad/$\mu$s) is much larger than any other scale. However, if they are further apart (say, 16 $\mu$m), the interaction is negligible (e.g. 0.32 rad/$\mu$s). This is the origin of the so-called **Rydberg Blockade**: within a certain radius, called the blockade radius, the low-energy subspace of time-dependent dynamics excludes the doubly excited Rydberg state. This Rydberg blockade mechanism is shown in Fig. 1.8.

Crucially, this strong interaction makes the dynamics insensitive to its exact value. If the energy of the doubly excited state is much larger than the Rabi drive $\Omega$ or detuning $\Delta$ (order GHz vs. MHz), it can be adiabatically eliminated from the effective dynamics. Algorithms and protocols can take advantage of this by approximating the doubly interacting state as completely excluded from dynamics ($V \to \infty$), which means that the mechanism for entanglement is robustly encoded not in specific values of the *interactions*, but instead in the *structure of the Hilbert space*.

# Engines on: a full cycle of the Aquila processor

These individual pieces are assembled in a sequence of well-timed and orchestrated laser pulses into a single measurement cycle, plotted in Fig. 1.9. Because the measurement process is destructive and based on removing atoms from traps, the entire array must be reloaded for every measurement shot. The reloading process is relatively slow in comparison to the ~10 $\mu$s time to execute the quantum computation; Aquila's shot rate is typically < 10 samples per second today. However, future systems will integrate methods that do not require complete reloading of the array every measurement cycle, which can feasibly increase the shot rate by orders of magnitude.

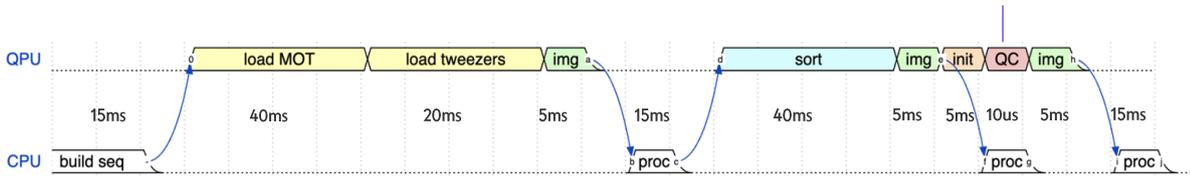

Figure 1.9. A full cycle of the Aquila processor. First, the magneto-optical trap (MOT) is loaded and then the static traps are loaded from the atoms in the MOT. Next, the occupancy of every randomly filled trap is imaged (img) and processed (proc), and the dynamic laser tweezers sort the array into the user-specified configuration. Another image is taken to determine the success of the sorting and is returned as the pre_sequence data key. Then, the quantum computation (QC) is executed on a fast $\mu$s time scale. Finally, the traps are turned back on, pushing away the Rydberg state and trapping the ground state to perform a measurement. The atom occupancy is imaged and returned as the post_sequence data key, which is interpreted as the bitstring measurement in the Z basis. Finally, the atoms are released back into the vacuum chamber and the cycle repeats, up to 10 times per second.

# 1.3. The Rydberg Hamiltonian

The final analog quantum dynamics are generated by a combination of the single-atom interaction between the laser field and the atom electronic state driving transitions between ground and Rydberg state, plus two-atom interactions between adjacent atoms shifting the energy of the state by a state-dependent Van der Waals interaction.

> **Definition of the Rydberg Hamiltonian**
>
> The evolution of the state is described by a time-dependent unitary generated by $H(t)$ evolved for a time T.
>
> $$|\psi\rangle = \mathcal{T}exp\left(-i\int_0^T H(t)\,dt\right)|0\rangle,$$
>
> where we choose units of $\hbar = 1$. the Hamiltonian is
>
> $$H(t) = \frac{\Omega(t)}{2}\sum_i e^{i\phi(t)}|g_i\rangle\langle r_i| + e^{-i\phi(t)}|r_i\rangle\langle g_i| - \Delta(t)\sum_i \hat{n}_i + \sum_{i<j}\frac{C_6}{|\vec{x}_i - \vec{x}_j|^6}\hat{n}_i\hat{n}_j.$$
>
> $|r_i\rangle \equiv |1\rangle$ is the representation of the Rydberg state on the ith qubit, and $|g_i\rangle \equiv |0\rangle$ is the representation of the ground state. $\hat{n}_i = |r_i\rangle\langle r_i|$ counts Rydberg excitations, and measurements are restricted to the logical Z basis only. There are four types of control parameters that, when specified, define a quantum program.
>
> $\Omega(t)$     The Rabi drive amplitude. This sets the frequency at which each qubit oscillates between ground and Rydberg state in the absence of interactions.
>
> $\phi(t)$     The phase of the Rabi drive. This sets the direction on the Bloch sphere around which the qubit is driven.
>
> $\Delta(t)$     Detuning. This sets how off resonant the global Rabi drive is.
>
> $\vec{x}_i$     The position of each atom in the array. This sets the Rydberg-Rydberg interaction strength between each qubit.
>
> This whitepaper uses units of radians per microsecond and micrometers. This contrasts with units of MHz by a factor of $2\pi$, which is the characteristic single-qubit oscillation frequency. Amazon Braket uses units of radians per second and meters, which differs from the conventions here by a factor of $10^6$.

## 1.4. Dominant sources of error

While quantum dynamics are extremely good due to the high stability and coherence of the lasers and atoms, as the evolution time increases, various noise sources reduce the fidelity of the state. There are several major noise sources which may contribute:

> **Best Practices tip:**
> When implementing dynamics, it is important to keep the protocol as short as possible. If possible, choose the maximum Rabi drive $\Omega$ to minimize the time given fixed pulse area $\Omega t$.

**Laser noise:**
While Aquila uses extremely stable lasers, they still suffer from some level of phase and amplitude noise. This causes a coherent shot-to-shot variance and time-dependent noise in the Rabi frequency $\Omega$ and detuning $\Delta$, which causes the shot-averaged expectation values to settle towards the time average.

**Atom motion:**
Each atom is held in an optical trap and laser cooled to extremely cold temperatures of order $\mu K$. However, there is some level of thermal motion in the traps, which causes a coherent shot-to-shot variance in the detuning $\Delta$ due to the Doppler shift of the laser frequency from each atom's velocity. This effect causes the shot-averaged expectation value to settle towards the time average and is most sensitive when the detuning and Rabi frequency are similar. Additionally, interactions may be sensitive to thermal position variations, which cause the atom to be at slightly varying distances from its neighbors and thus have varying Rydberg interactions $V_{ij}$.

**State decoherence and scattering:**
A major source of error is due to incoherent decay of the atomic energy levels back to the ground state. Aquila uses a two-photon transition via an off-resonant intermediate state to drive the transition between ground and Rydberg states. This intermediate state, as well as the excited Rydberg state, decay incoherently back to the ground state or other intermediate state in a lossy process. The Rydberg state is technically always hybridized with the intermediate state during evolution, which may cause loss even if the Rabi drive is turned off. This effect is especially noticeable for the Ramsey protocol, where the Rabi drive is turned off for a variable time to track dephasing noise.

**Inhomogeneity:**
Due to imperfect holography of the Rydberg lasers, the Rabi frequency and detuning may be slightly different across the 2D array.

**Measurement:**
One leading cause of error occurs from state measurement; because the state readout is given by the presence or absence of an atom, an imperfect re-trapping causes an incorrect readout. This may be either a) incorrectly reading out a ground state (atom presence) as a Rydberg state, due to atom loss, or b) incorrectly reading out a Rydberg state (atom absence) as a ground state, due to imperfect Rydberg anti-trapping.

## 1.5. Datasheet of Aquila capabilities and performance metrics

Listed below are several key restrictions on allowed programs implementable on Amazon Braket, as well as a brief explanation as to what limits the value physically. These numbers were chosen to be well within the performance limits of Aquila to guarantee expected behavior and can be relaxed if needed. For example, certain experiments shown in this document show coherent evolution out to 10 $\mu s$ and geometries with a vertical height of 115 $\mu m$. Interested users are encouraged to inquire about "premium access" mode with these and other extended capabilities. Note that there are several other technical restrictions on valid programs; for more details, see the full [Amazon Braket specifications](#).

| Restriction | Value | Reason |
| --- | --- | --- |
| Maximum total number of filled and unfilled user-defined sites | 256 | Limited laser power for tweezers that generate the user-defined and reservoir sites. |
| Maximum number of qubits, i.e. maximum number of filled sites | 256 | Finite filling rates when loading the atom array and limited laser power for laser tweezers. |
| Maximum site pattern width<br>Maximum site pattern height | 75 $\mu m$<br>76 $\mu m$ | Position-dependent tweezer power efficiency due to the holography method used. |
| Minimum distance between user-defined sites | 4 $\mu m$ | Optical resolution of the imaging system and of the holography method used. |
| Minimum vertical spacing between rows | 4 $\mu m$ | Packing geometry of the reservoir sites. |
| Maximum Rydberg Rabi frequency | $15.8 \frac{rad}{\mu s}$ | Limited delivered laser power for driving the ground-Rydberg transition. |
| Rabi slew rate | $\left|\frac{d\Omega}{dt}\right| \leq 250 \frac{rad}{\mu s^2}$ | Response speed of the acousto-optic device responsible for modulating the Rydberg drive laser. |
| Detuning range | $|\Delta| \leq 125 \frac{rad}{\mu s}$ | Electronic bandwidth of the acousto-optic device responsible for modulating the Rydberg drive laser. |
| Maximum duration of user-defined evolution | 4 $\mu s$ | Approximate timescale of coherent evolution. |

| Name | Value | Explanation |
|---|---|---|
| **State preparation, measurement, and Hamiltonian errors** | | |
| $\delta_x, \delta_y$ | $0.050\ \mu m$ | Systematic, pattern-dependent error between specified and actual lattice site positions. |
| $\sigma_x, \sigma_y$ | $0.200\ \mu m$ | Random error in the qubit positions during coherent evolution as a result of thermal atom motion. |
| $\langle\delta\Omega^2\rangle^{1/2}/\langle\Omega\rangle$ | 0.02 | RMS relative Rabi frequency inhomogeneity over the user region. |
| $\langle\delta\Delta^2\rangle^{1/2}$ | $0.37\ \dfrac{rad}{\mu sec}$ | RMS detuning inhomogeneity over the user region. |
| $\delta_\Delta$ | $0.63\ \dfrac{rad}{\mu sec}$ | Systematic error in global detuning from specified value. |
| $\langle\delta\Delta^2\rangle^{1/2}$ | $0.18\ \dfrac{rad}{\mu sec}$ | RMS shot-to-shot variance in the detuning. |
| $\langle\delta\Omega^2\rangle^{1/2}/\langle\Omega\rangle$ | 0.008 | RMS relative shot-to-shot variance in the Rabi frequency. |
| $\epsilon_{fill}$ | 0.007 | Typical probability of failing to occupy a site specified by user as 'filled'. These probabilities are dependent on the pattern and site position within the pattern. |
| $\epsilon_{det-fn}$ | 0.01 | Probability of false-negative atom detection error (mis-detecting a filled site as empty). |
| $\epsilon_{det-fp}$ | 0.01 | Probability of false-positive atom detection error (mis-detecting an empty site as filled). |
| $\epsilon_{det-gnd}$ | 0.01 | Probability of mis-detecting a ground-state atom as a Rydberg |
| $\epsilon_{det-ryd}$ | 0.08 | Probability of mis-detecting a Rydberg atom as a ground-state atom |
| **Ground-Rydberg qubit coherence** | | |
| $T_2^*$ | $5.8\ \mu sec$ | Qubit dephasing time without drive from individual non-interacting qubits, including coherent and incoherent processes, as measured by a Ramsey protocol (see example 1). |
| $T_2^{echo}$ | $11.4\ \mu sec$ | Qubit dephasing time without drive from individual non-interacting qubits from incoherent processes only, as measured by a spin-echo dynamical decoupling protocol (see example 1). |
| $T_2^{Rabi}$ | $7.5\ \mu sec$ | Driven qubit decoherence under maximum Rabi frequency from individual non-interacting qubits, including coherent and incoherent processes, as measured by Rabi oscillations (see example 1). |
| $T_2^{blockaded-Rabi}$ | $8.9\ \mu sec$ | Driven qubit decoherence under maximum Rabi frequency for an isolated pair of mutually blockaded qubits, including coherent and incoherent processes as measured by Rabi oscillations (see example 2). |
| **Application benchmark metrics** | | |
| 1D correlation length | 3.6 sites | The correlation length of an adiabatically prepared $Z_2$ state (see example 3) |
| 2D correlation length | 5.7 sites | The correlation length of an adiabatically prepared checkerboard state (see example 3) |

# 2. Example 1: Single-qubit dynamics

Now that we know the basics of how Aquila works, let's try it out! The second half of this document will go over various examples and demonstrations of increasing complexity. This first example is the quantum version of "Hello World": **single-qubit dynamics**, without interactions or entanglement between atoms. The associated Jupyter notebook is here. For one qubit, there are only two states, $|0\rangle$ and $|1\rangle$, and the wavefunction of the qubit can be described by two angles $\theta$ and $\phi$:

$$|\psi\rangle = \cos(\theta)\,|0\rangle + \sin(\theta)\,e^{i\phi}\,|1\rangle.$$

In the *ground-Rydberg qubit* used in Aquila, the logical 0 state is associated with the ground state $|0\rangle \equiv |g\rangle$, and the logical 1 state is associated with the excited Rydberg state $|1\rangle \equiv |r\rangle$. Depending on the context, these conventions will be used interchangeably. Any single-qubit wavefunction can be prepared in several ways. One canonical way is via two parameterized single-qubit gates that rotate the $|0\rangle$ state. The first $Y(\theta)$ gate rotates the state into an X superposition state, then the second $Z(\phi)$ phase gate adds the phase to the $|1\rangle$ state. The second way to prepare a single-qubit wavefunction is via a time-dependent single-qubit Hamiltonian

$$H(t) = \frac{1}{2}(\Omega(t)|g\rangle\langle r| + \Omega^*(t)|r\rangle\langle g|) - \Delta(t)\,|r\rangle\langle r|,$$

where $\Omega(t)$ is a complex value corresponding to the laser amplitude and phase, and $\Delta(t)$ is a real value corresponding to the detuning of the laser from resonance between the ground and Rydberg state. By choosing $\Omega(t)$ and $\Delta(t)$, one can prepare any superposition state: the total integrated $\Omega t$ can be related to the angle $\theta$, and the total integrated $\Delta t$ can be related to the angle $\phi$.

An intuitive way to think about single-qubit evolution is using the **Bloch sphere**. The amplitude of the wavefunction is normalized to 1, so the 2D complex vector can be equivalently mapped to a point on a unit sphere, where the position is denoted by latitude and longitude $(\theta, \phi)$, as shown by the red point in Fig. 2.1. The Hamiltonian can similarly be denoted on the unit sphere; given a conversion of parameters $\Omega$ and $\Delta$ to effective magnetic fields $h_x, h_y, h_z$, the Hamiltonian can be written as a sum of Pauli terms $H = \vec{h} \cdot \hat{\sigma}$, and is a vector of norm $|H|$ pointing at $\vec{h}$ on the unit sphere, as shown by the purple arrow in Fig. 2.1. Time evolution of the Schrodinger equation is equivalent to a precession of the unit vector around the vector $\vec{h}$, as shown by the green arrow in Fig. 2.1. In this way, any single-qubit state can be prepared by the appropriate time-dependent vector $\vec{h}$ and evolution time $t$.

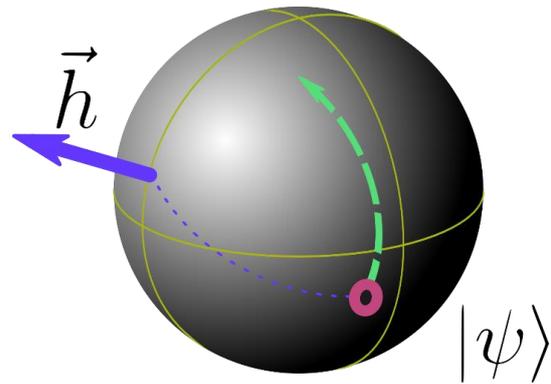

Figure 2.1. Single-qubit dynamics on the Bloch sphere. The 2D complex vector $|\psi\rangle$ can be represented as a point on a sphere (red). The Hamiltonian can be represented as a vector $\vec{h}$ (purple), and time evolution with respect to $H$ is precession around the effective magnetic field (dashed).

## 2.1. Rabi oscillations

**Rabi oscillations** are a canonical demonstration of time-dependent behavior, where Ω and Δ are held fixed, and the total time $t$ is varied, driving the atom between the states $|0\rangle$ and $|1\rangle$. If the detuning $\Delta = 0$, the state oscillates sinusoidally between the initial state $|0\rangle$ and the excited state $|1\rangle$; for nonzero detuning, the state oscillates between $|0\rangle$ and some superposition state. Some examples of these Rabi oscillations are shown in the top half of Fig. 2.2. Observe that, as expected by the black line of exact dynamics, the probability of measuring the $|0\rangle$ ground state oscillates between 0 and 1 for detuning $\Delta = 0$. Similarly, the probability of measuring the ground state oscillates between 0 and 50% for $\Delta = \Omega$, as expected.

**Best Practices tip:**
When doing few-atom dynamics, you may reduce the number of shots by parallelizing the same configuration multiple times across the array. This example uses a 4 × 4 grid of atoms spaced 25$\mu m$ apart (see Fig. 1.7).

Observe that short evolution times < 100ns are accessible due to the finite rise and fall time of the laser intensity, which manifests as a restriction on the valid quantum programs that can be run on Aquila.

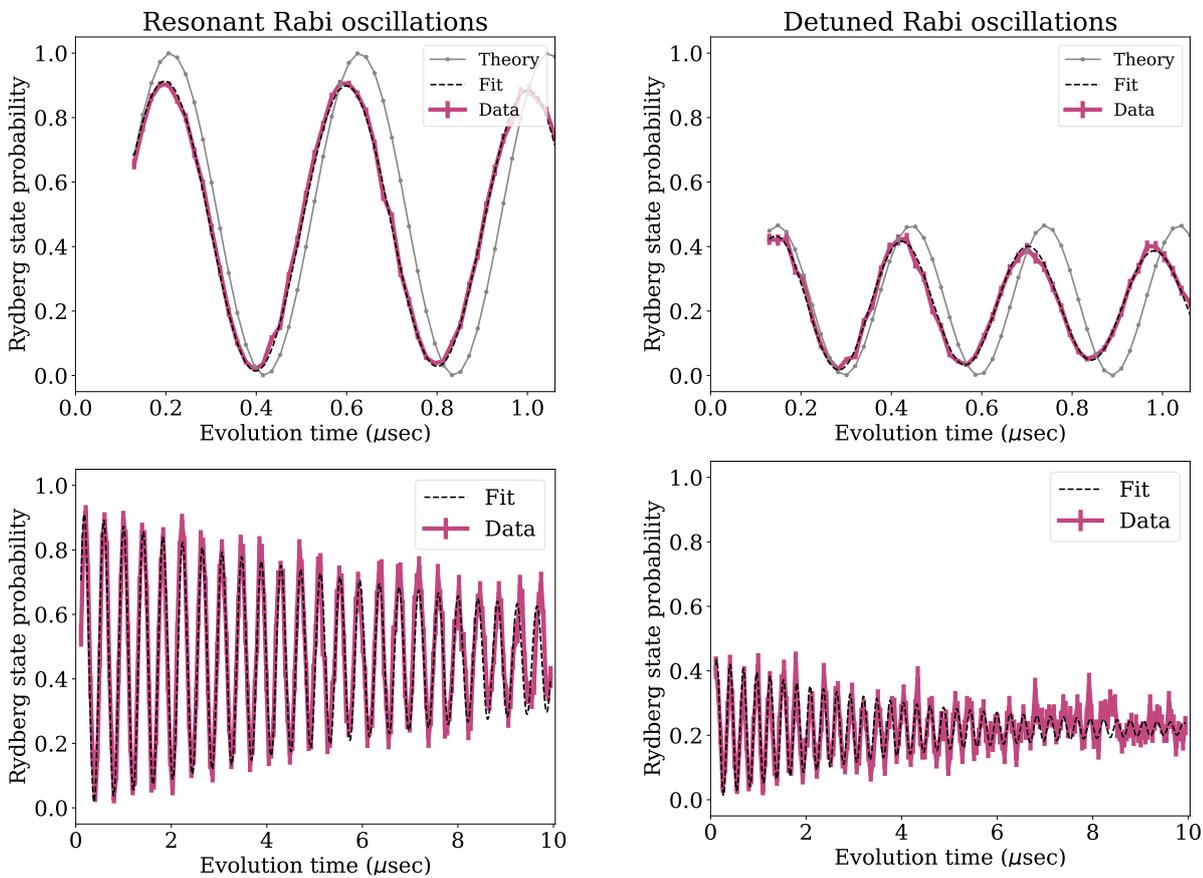

Figure 2.2. Single-qubit Rabi oscillations averaged over an ensemble of 16 atoms. The laser is turned on for a variable time, driving each qubit between the ground state and Rydberg state sinusoidally. Left, the atoms are driven resonantly with $\Omega = 15$ rad/$\mu s$ and $\Delta = 0$. Right, the atoms are driven off-resonantly with $\Omega = \Delta = 15$ rad/$\mu s$. Top plots show short-time data, while bottom plots data over a longer evolution time and single atoms. Gray is the theoretically expected value, which differs slightly from data due to mis-calibrations in resonance and decoherence within tolerance. Black line is a best-fit curve. Each point averages over 40 measurements parallelized across 16 atoms, for a total of 640 measurements per point. Note that the maximum time is greater than the 4 $\mu s$ restriction on Amazon Braket and is available in "premium access" mode.

While short-time dynamics are extremely good, evolution for longer times begins to suffer from the effects of noise and decoherence. Some example Rabi oscillations are shown in the bottom half of Fig. 2.2, which are the same as top except running through a longer evolution time. For longer times, the state begins to decay due to coherent and incoherent noise. We can fit the sinusoidal oscillations to a heuristic fit

$$Z(t) = A\sin(\Omega t + \phi)e^{-t/\tau} + B$$

which is a good approximate model of Rabi oscillations under incoherent decay. The value $A$ is related to the contrast $\Omega^2/(\Omega^2 + \Delta^2)$ multiplied by the measurement error rate, and the constant $\tau$ is the coherence time of the Rabi oscillations. The total "analog gate depth" can be seen as $\Omega\tau/\pi$, which is the number of $\pi$ flops before the system begins to decohere. For on-resonant oscillations, we fit a time constant $\tau \approx 3.6 \, \mu s$ when averaged across 16 sites, which includes dephasing from slightly inhomogeneous Rabi drive; each atom individually has an average time constant $\tau \approx 7.5 \mu s$. This value corresponds to an effective gate depth of $\Omega\tau/\pi \approx 32$. For off-resonant oscillations, which are maximally sensitive to phase fluctuations, we fit a time constant $\tau \approx 2.6 \, \mu s$; each atom individually has an average time constant $\tau \approx 3.6 \, \mu s$.

> **Best Practices tip:**
> When parallelizing across the array, be aware that the detuning and Rabi frequency may be slightly inhomogeneous across the array. Typically, this effect is most pronounced in the X dimension.

## 2.2. Time-dependent protocols

**One strength of analog mode computation is the capacity to choose arbitrary time-dependent waveforms** for the detuning and Rabi drive. Let us demonstrate this with two time-dependent protocols, as shown in Fig. 2.3. The first is a Ramsey protocol, which is used to calibrate the resonance condition of where to set $\Delta = 0$. This protocol occurs in three parts. In part one, the Rabi drive is turned on for a calibrated amount, so that the total integrated drive $\Omega\tau = \pi/2$ converting the state to a superposition $\frac{1}{\sqrt{2}}(|0\rangle + i|1\rangle)$. In part two, the Rabi drive is turned off, so that the state only accumulates a time-dependent phase on the Rydberg state $\frac{1}{\sqrt{2}}(|0\rangle + ie^{i\phi}|1\rangle)$. Finally, the third part turns on again to execute the same $\pi/2$ rotation around X; depending on the accumulated phase $\phi = \Delta\tau$, the state is returned to the ground state or pushed completely to the Rydberg state. As expected, the state decays as a function of time due to noise and decoherence. There are coherent effects, such as variance in detuning from laser and Doppler effects, as well as incoherent effects due to the hybridization and decay with the intermediate state. Using this and similar methods, we find a coherence time $T_2^* \approx 5.5\mu s$ if one averages across the ensemble of atoms in the array, or $T_2^* \approx 5.8 \, \mu s$ if one does not do the averaging. This coherence time is a measure of the phase stability of the coherent evolution.

A second non-quantitative protocol is chosen to showcase the arbitrary waveform capabilities of Aquila's analog mode and is shown in Fig. 2.3 Right. In this protocol, the Rabi drive is held constant $\Omega = 15$ rad/$\mu s$, while the detuning is varied as a function of time as $\Delta = \Delta_0 \sin(\omega t)$, where $\Delta_0 = \omega = 15$ rad/$\mu s$. This is a Floquet evolution in the non-perturbative regime where all parameters are the same order. As the qubit evolves along the protocol, the state changes coherently to nontrivially explore the Bloch sphere. As can be seen, the theoretical expectation and experimental data line up relatively well, though performance is otherwise qualitative.

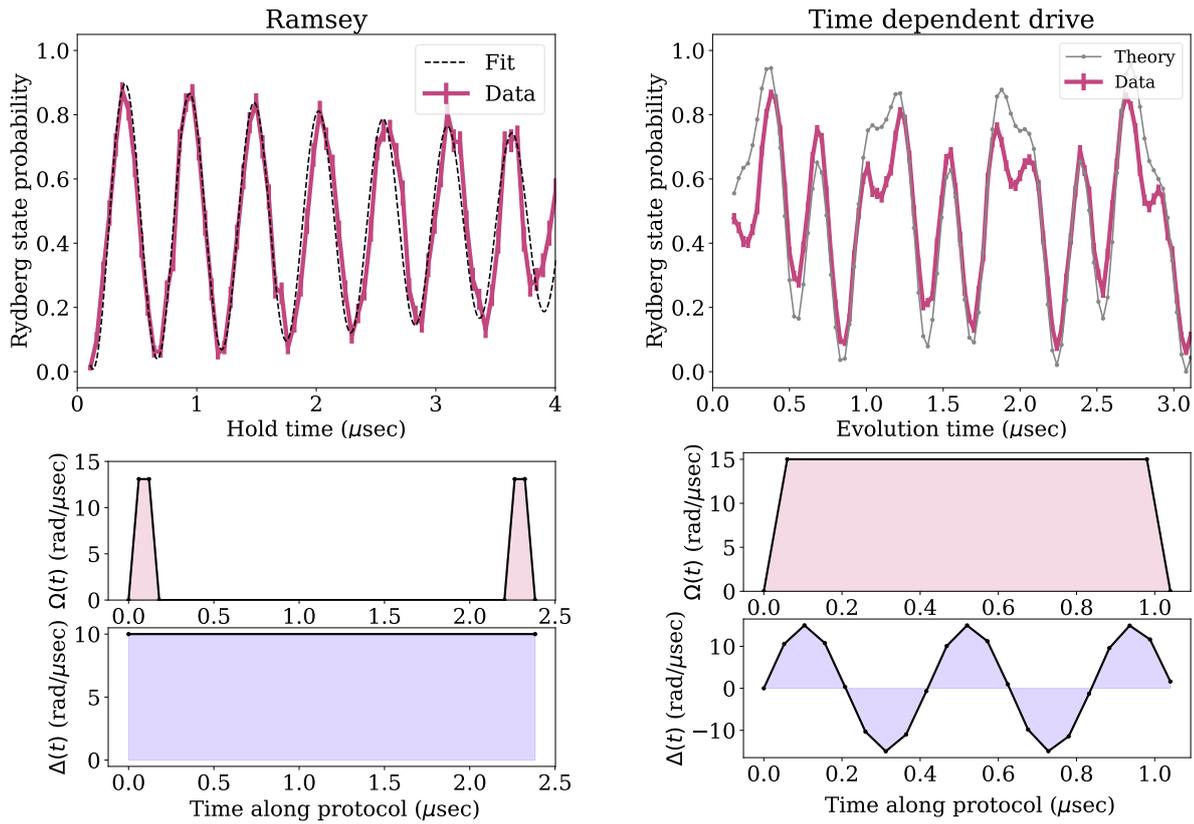

Figure 2.3 Time-dependent protocols. Left is a Ramsey protocol, where the state is rotated into a coherent superposition, held for some time, and then rotated back. The hold time is varied to get a sinusoidal behavior. Right is a time-dependent protocol inspired by a Floquet drive. For different increments of time, the detuning is driven sinusoidally as $\Delta = 15\sin(15t)$ (units rad/μsec).

## 2.3. Dynamical decoupling protocols

A time-dependent protocol can also take advantage of the phase control of the laser to drive rotations on the Bloch sphere in different directions. One simple example that uses phase control is the spin-echo protocol, which is a dynamical decoupling scheme that removes any coherent phase drift to first order. The protocol is a sequence of three single-qubit rotations, with a variable wait time in between, and the analog version is shown in Fig. 2.4 Right. Similar to the previous example, the first gate pulse implements a $\pi/2$ rotation around the X axis, preparing a $\frac{1}{\sqrt{2}}(|0\rangle + i|1\rangle)$ state from the initial $|0\rangle$ state. Then, the qubit evolves freely under noise from the environment, accumulating some phase $\Delta_{env} t$. The second gate pulse implements a $\pi$ rotation around the Y axis by shifting the laser phase. If there is no phase accumulation, nothing happens as the state is an eigenstate of the pulse; any phase accumulation is swapped in sign $\Delta_{env} t \rightarrow -\Delta_{env} t$. Next, the qubit evolves freely under noise again, which cancels out the phase accumulation from the environment to first order, with an additional detuning offset inducing an overall phase accumulation and a sinusoidal response. Finally, a $-\pi/2$ pulse rotates the state back to $|0\rangle$ if there is no phase accumulation. Data for varying hold time $t$ is shown in the left half of Fig. 2.4. We find a coherence time of $T_2^* \approx$ 11.4 $\mu s$, which is inherently larger than the Ramsey protocol due to cancelation of first-order phase decoherence.

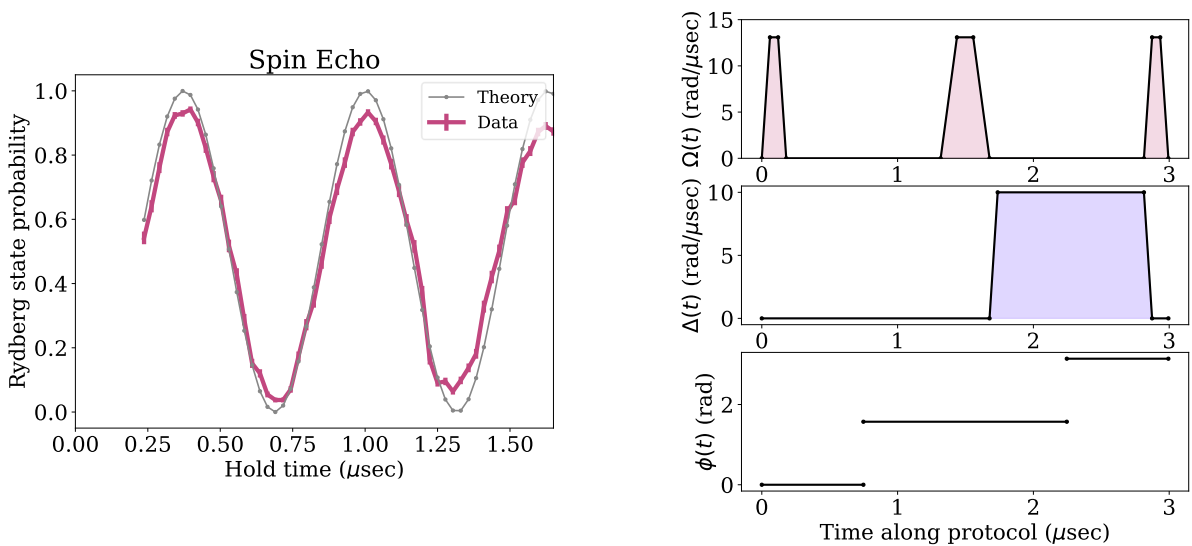

Figure 2.4 A spin-echo protocol. The state is rotated into a coherent superposition, and environmental phase accumulation is canceled out at first order by a Y rotation mid-sequence. The hold time is varied to get a sinusoidal dependence; the total evolution time is double the hold time, plus the time to execute the single-qubit flip terms.

# 3. Example 2: Many-qubit dynamics

In this example, we will demonstrate how interactions between atoms can be used to generate entanglement and correlations. Given multiple atoms indexed by $i$ at positions $\vec{r}_i$, the Rydberg-Rydberg interaction goes as

$$V_{ij}|r_i r_j\rangle\langle r_i r_j| \quad \text{where} \quad V_{ij} = \frac{C_6}{|\vec{r}_i - \vec{r}_j|^6}$$

Here, $C_6 = 5{,}420{,}503\ \mu m^6 \text{rad}/\mu s$ is the Rydberg interaction coefficient for the $70S$ state; as a characteristic scale, for $R = 8\mu m$, $V_{ij} \approx 20$ rad/$\mu$sec. Due to the large exponent $R^{-6}$, there is a rapid falloff in the interaction as a function of distance. If two atoms are close together (e.g., $4\ \mu m$), the energy of the doubly excited state (e.g., 1,320 rad/$\mu sec$) is much larger than any other scale. However, if they are further apart (e.g., $16\mu m$), the interaction is negligible (e.g., 0.32 rad/$\mu sec$). This is the origin of the so-called **Rydberg blockade**: within a certain radius, called the **blockade radius**, the low-energy subspace of time-dependent dynamics excludes the doubly excited Rydberg state. This subspace is insensitive to the variance of position of each atom, as it only requires that the scale be much larger than the Rabi frequency Ω and detuning Δ. For more details, see this page of Bloqade documentation, and the associated Jupyter notebook is here.

> **The Rydberg blockade**
>
> *Suppose two atoms at some distance R from each other each at some detuning $\Delta > 0$ and Rabi drive $\Omega = 0$. If $\Delta > V_{ij}$, it is energetically favorable for the ground state to be the doubly excited Rydberg state $|\emptyset\rangle = |rr\rangle$. However, if $\Delta < V_{ij}$, the atoms are blockaded and it is energetically favorable for the ground state to only have one Rydberg state, and so will be an entangled superposition $|\emptyset\rangle = \frac{1}{\sqrt{2}}|gr\rangle \pm \frac{1}{\sqrt{2}}|rg\rangle$. The radius at which $\Delta = V_{ij}$, which is the turnover to where it is energetically preferred to blockade one of the atoms, is called the **static blockade radius** $R_b = (C_6 / \Delta)^{1/6}$.*
>
> *Alternatively, suppose two atoms at some distance R from each other at some nonzero Rabi drive $\Omega > 0$ and zero detuning $\Delta = 0$. If the atoms are far apart and $\Omega \gg V_{ij}$, the atoms undergo coherent flips independently. If $\Omega \ll V_{ij}$, it is energetically prohibited for the drive to flip both atoms into the Rydberg state, and the doubly excited Rydberg state is blockaded. The radius at which $\Omega = V_{ij}$, where the energy scale of flipping an atom from the ground to the excited state is equal to the interaction, is called the **dynamic blockade radius** $R_b = (C_6 / \Omega)^{1/6}$.*
>
> *If both detuning and Rabi drive are nonzero, then the two effects can be combined as a characteristic energy scale to define the **blockade radius** $R_b = \left(C_6 / \sqrt{\Omega^2 + \Delta^2}\right)^{1/6}$. Note that depending on context, the blockade radius may equivalently mean either the static or dynamic blockade radius.*

## 3.1. Adiabatic state preparation and the Rydberg blockade

One demonstration of this blockade radius is adiabatic state preparation. Under this algorithm, Hamiltonian parameters are slowly ramped from some initial Hamiltonian to some target Hamiltonian. If slow enough under the adiabatic condition, an initial state that is the ground state of the initial Hamiltonian evolves to become the ground state of the target Hamiltonian. In this example and further examples to come, this protocol is shown in Fig. 3.1. The state is initialized in the ground state $|00\rangle$

> **Best Practices tip:**
> Adiabatic protocols are a flexible and robust mode for analog computation, due to a relative insensitivity to phase, amplitude, and position noise. Consider them when designing analog algorithms!

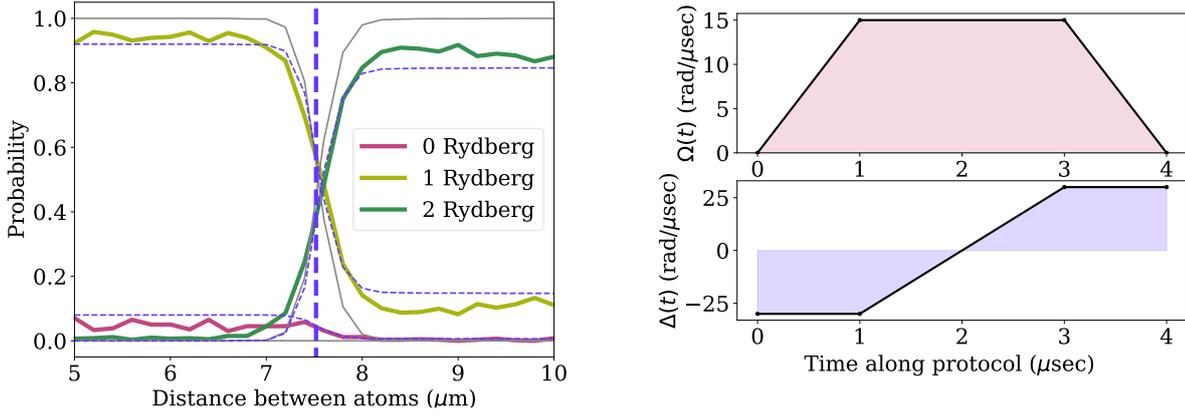

Figure 3.1. Probing the Rydberg blockade radius with an adiabatic state preparation. The ground states of two atoms with a variable distance are prepared adiabatically. Below the blockade radius (thick dashed line), the ground state has a single excitation (yellow); above the radius there are two excitations (green). Right plots the adiabatic protocol to prepare the ground state, which is typical of the piecewise linear protocols used in this whitepaper. The final detuning of 30 rad/μs corresponds to the dashed line blockade radius of $R_b \approx 7.51 \mu m$. Theory (grey; purple dashed lines include 8% measurement error) shows the noise-free prediction, which still includes some finite-time diabatic errors.

with the Hamiltonian $\Omega = 0$, $\Delta < 0$. Then, $\Omega$ is slowly ramped on; $\Delta$ is ramped from negative to positive; then $\Omega$ is slowly ramped off.

The results of this adiabatic protocol are shown in Fig. 3.1, where different programs scan different distances between the two atoms. Depending on the distance, the final state will have either one or two atoms in the Rydberg state, with the transition occurring at the static blockade radius. Note that even though this transition is perfectly sharp in the adiabatic limit, finite-time diabatic errors as well as noise and decoherence broaden this transition.

## 3.2. Rabi frequency enhancement

Another example of Rydberg blockade physics is Rabi frequency enhancement. Given $N$ atoms all within the blockade radius, there are only two relevant low-energy states: the ground state and a symmetric superposition (also known as a W state) of a single Rydberg excitation; other superpositions are not coupled to the ground state. The matrix element between those two states can be computed as

$$\frac{\langle 1000 \ldots | + \langle 0100 \ldots | + \langle 0010 \ldots | + \cdots}{\sqrt{N}} H |0000 \ldots \rangle = \frac{\Omega}{2}\sqrt{N}$$

In other words, the effective Rabi frequency of the two-level system is enhanced by a factor of $\sqrt{N}$. This effect is shown in Fig. 3.2 for different numbers of atoms within the blockade radius. Crucially, these dynamics are insensitive to the exact value of $V_{ij}$, as the only requirement on these blockaded dynamics is that the distance between any two pairs of atoms $i$ and $j$ must be much less than the blockade radius. The $\sqrt{N}$ scaling is clear in Fig. 3.2 bottom-right in comparison to a fit-free curve. We find single-cluster $T_2$ times of 8.9, 6.9, 6.6 $\mu s$ for $N = 2, 3, 4$ respectively. This effect is a clear demonstration of entangling dynamics, as the system dynamically prepares an $N$-body W state.

For $N = 2, 3, 4$, the maximum distance between atoms is $5.65 \mu m$, which is the diagonal of the square of side length $4 \mu m$ and has an interaction strength $V_{ij} \approx 165 \frac{rad}{\mu sec} \gg 15.0 \frac{rad}{\mu sec} = \Omega$. For $N = 7$, the maximum distance between atoms is $8.94 \mu m$, which is the diagonal of the hexagon and has an interaction strength $V_{ij} \approx 10.58 \frac{rad}{\mu sec} \gg 5.00 \frac{rad}{\mu sec} = \Omega$. Observe that for these larger number of atoms, the Rabi frequency does not have as much of a separation of scales, which causes some non-blockaded dynamics as is evident from Fig. 3.2.

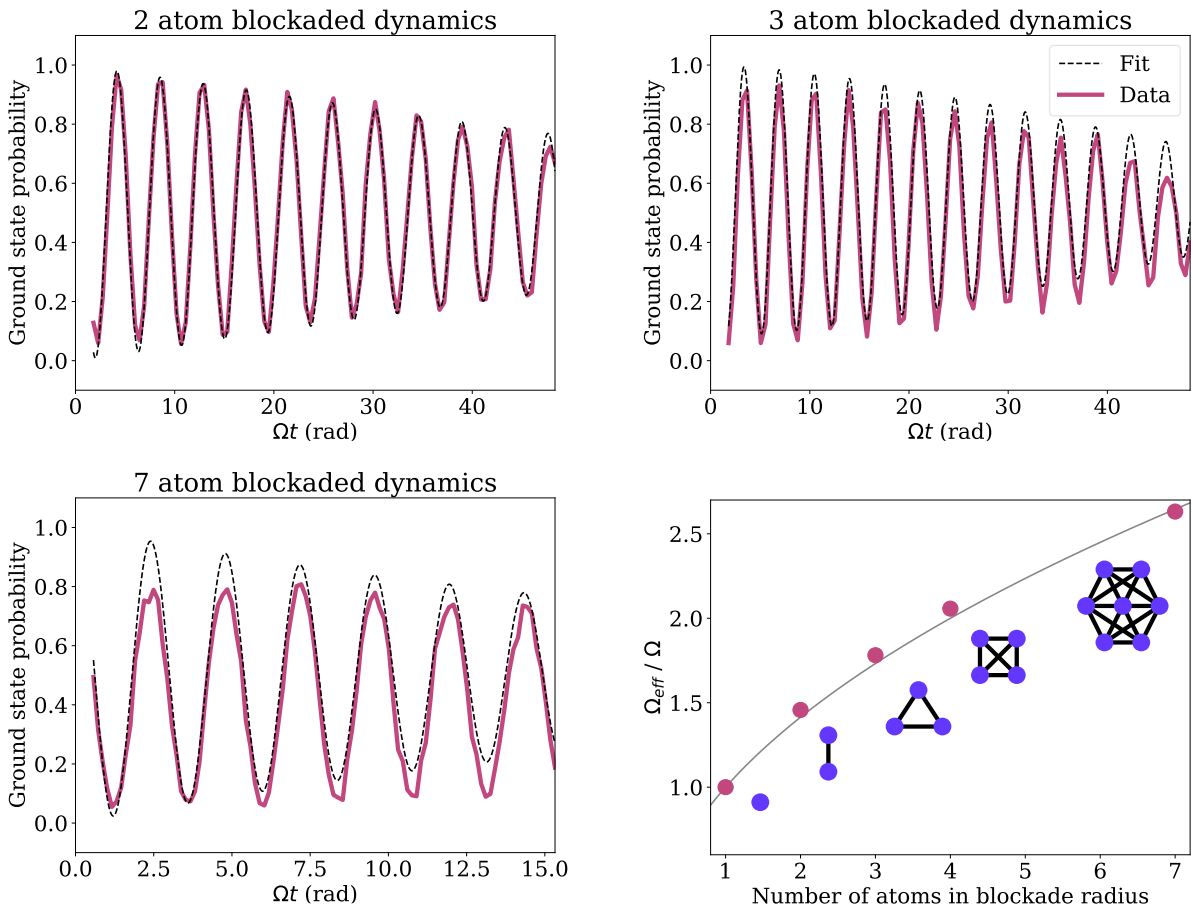

Figure 3.2 Rabi oscillations of $N$ atoms under the Rydberg blockade. There is a $\sqrt{N}$ enhancement of the effective Rabi frequency of the two-level system. 1, 2, 3 and 4 atom dynamics have a single-atom Rabi frequency of $\Omega = 15 \frac{rad}{\mu sec}$, while 7 atom dynamics have a single-atom Rabi frequency of $\Omega = 5.0 \frac{rad}{\mu sec}$ to stay within the blockade radius. Bottom right plots the best-fit effective Rabi frequency for different sizes, scaled by the single-atom Rabi frequency. There is a $\sqrt{N}$ enhancement (fit-free black line) as expected.

**Best Practices tip:**
To preserve blockaded dynamics, it is important that there is a separation of scales between the interaction and $\Omega$ by designing atom spacings that are either deep in or far from the blockade radius.

## 3.3. Levine-Pichler gate analogues

Underlying every digital gate execution is a sequence of analog waveforms that implement the gate. A demonstration of analog control is to execute the **Levine-Pichler (LP) gate**, which is a neutral-atom controlled-Z phase gate with hyperfine qubits mediated by the Rydberg state [Levine2019]. By taking advantage of the $\sqrt{2}$ enhancement of the Rabi frequency, the gate accumulates a different dynamical phase depending on the initial hyperfine state of two atoms using a global Rydberg drive between the $|1\rangle$ state and the $|r\rangle$ state and a phase jump. For more details, we refer to [Levine2019] or Fig. 3.3 as reproduced from that paper.

While the LP gate requires two hyperfine ground states to encode a qubit, it is possible to reproduce the dynamics of the Rydberg state population with only the ground-Rydberg qubit. This can be done by implementing the LP pulse protocol on either one or two atoms. If there are two atoms, this is analogous to dynamics of the $|11\rangle$ initial state, which has the $\sqrt{2}$ Rabi enhancement. If there is one atom, this is analogous to dynamics of the $|01\rangle$ or $|10\rangle$ initial state.

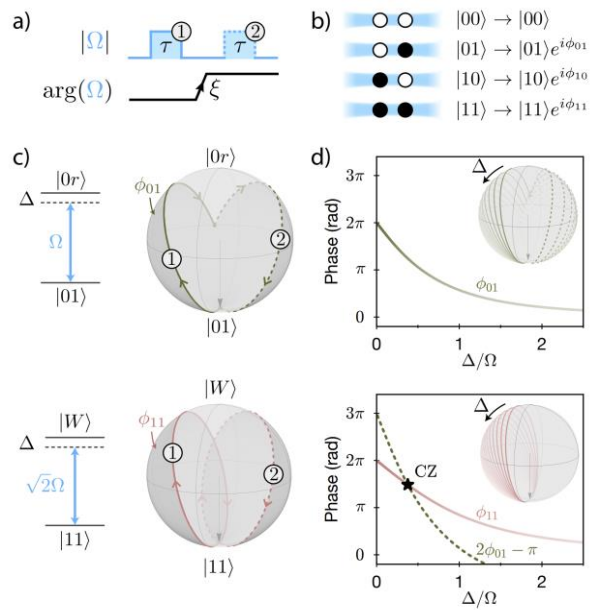

Figure 3.3 A diagram of the Levine-Pichler gate. By choosing a particular phase jump $\xi$, a global field implements a CZ gate using a state-dependent dynamical phase. Reproduced from [Levine2019].

Trivially, zero atoms are analogous to the $|00\rangle$ state, as there is no Rydberg coupling between the $|0\rangle$ and $|r\rangle$ state. This does not implement the LP gate, as the conditional dynamical phase is instead a global phase of the ground-Rydberg qubits. However, it does act as an upper bound on the fidelity by observing the probability that the atoms return to the ground state at the end of evolution.

Data from Aquila for the LP analogue is shown in Fig. 3.4. We find using a fixed, unoptimized protocol that the ground state probability for one and two atoms is 97.9% and 96.0%, respectively. This serves as an extremely weak upper bound on the fidelity of the LP gate, as the gate protocol requires that the state end with no density in the Rydberg state at the end of evolution.

**Best Practices tip:**
When executing a large jump in phase, it is best to not have the Rabi drive active. This is due to the particulars of the AOMs that drive the phase and amplitude of the Rabi drive.

It is important to emphasize that these return probabilities are not indicative of gate fidelities on future neutral-atom hardware. Aquila is optimized for analog-mode computation, where certain time-dependent characteristics are de-emphasized in favor of other analog characteristics. **There is no actual gate that is executed**, as Aquila does not have access to the two hyperfine states that represent a qubit. Additionally, there was no optimization done on the particulars of the protocol and was instead simply fixed by the theoretically optimized parameters given by [Levine2019]. Finally, the LP gate is in fact not an optimal gate in terms of total time or other metrics. More recent theory works [Jandura2022] find better time optimal gates. Nonetheless, these simple suboptimal protocols surpass the fidelity of work several years ago [Jau2016]. The implementation of the LP gate in 2019 found a fidelity of 97.4% [Levine2019], with a recent work [Evered2023] demonstrating gate fidelities upwards of 99.5%.

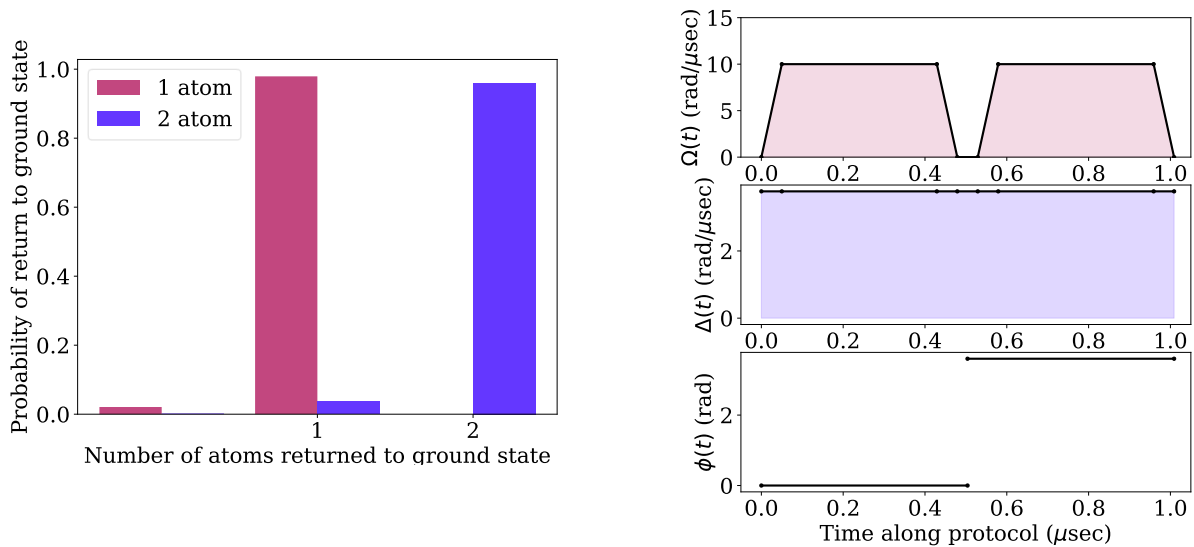

Figure 3.4 The LP gate analogue, which reproduces the gate pulses that implement a two-qubit CZ gate with hyperfine qubits. Top plots the probability of the atoms returning to the ground state for one or two atoms; this value is a weak proxy for the fidelity of the LP gate. We observe a 97.9% probability for 1 atom and 96.0% probability for 2 atoms. Right plots the protocol as a function of time. Observe that the Rabi drive is turned off when the phase is shifted.

## 3.4. Interacting non-equilibrium dynamics of two atoms

As a final non-quantitative example, let us demonstrate some interesting non-equilibrium dynamics. Instead of putting atoms well within the blockade radius, we position two atoms at 8.5 $\mu m$ from each other, so that the interaction strength $V_{ij} \approx 14.37 \frac{rad}{\mu s}$ is similar to the Rabi drive $\Omega = 15.0 \frac{rad}{\mu s}$. In this way, the time dynamics are in between two independent qubits ($V_{ij} \ll \Omega$) and two blockaded qubits ($V_{ij} \gg \Omega$), so that the probability of each state evolves non-trivially in time.

**Best practices tip:**
When designing protocols and algorithms, it is best to set atoms either deep within or far from the blockade radius; otherwise, the evolution may be sensitive to thermal position fluctuations.

Results for this protocol are shown in Fig. 3.5. At short times, the probability density follows that of the theoretical expectation, while at longer times the dynamics diverge from theory, due to single-atom decoherence as well as thermal fluctuations. On a shot-to-shot basis, the exact position of each atom fluctuates from the user-specified value by a small amount (typically on the order of 200 nm). When atoms are well within the blockade radius or far away, this small variance does not contribute much to error. However, a 200 nm variance at the blockade radius corresponds to a variance of order $4 \frac{rad}{\mu s}$, which dynamics are sensitive to.

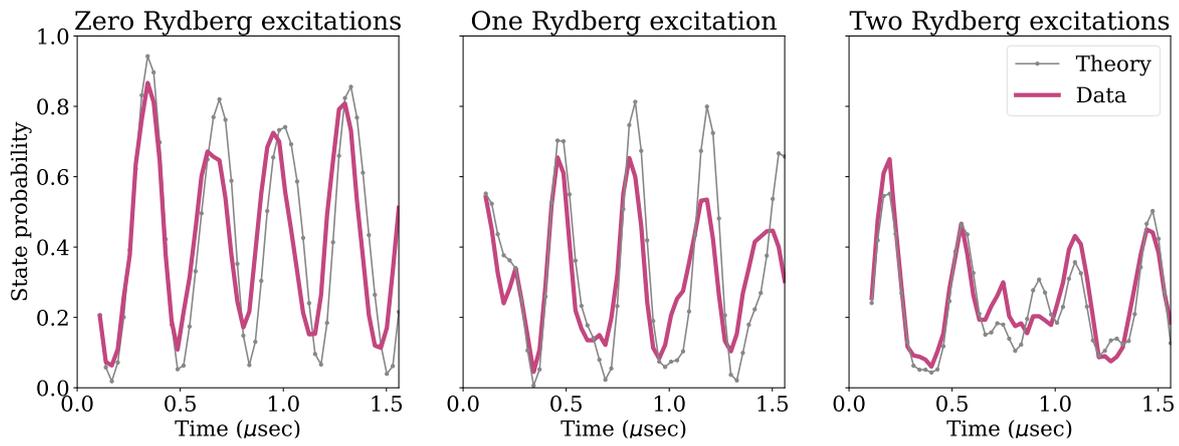

Figure 3.5 Minimally nontrivial dynamics of two atoms at the blockade radius. The Rabi frequency is turned on for a variable time for two atoms spaced 8.5 *μm* apart, and the number of Rydberg states is counted for $N = 40$ measurements per time step. Observe that at short times the theoretical prediction matches the experimental results, while at longer times they diverge due to single-atom decoherence effects as well as the thermal fluctuations of the positions of each atom.

# 4. Example 3: Many-body ordered phases

Now we see examples with few-atom interactions. This example extends to the case with many interacting atoms, for up to 256 qubits. Typically, the dynamics and quantum behavior of these systems extend well beyond the computational power of classical emulation, which typically requires exponential resources in the number of qubits ($2^{256}$). In this example, instead of studying quantum dynamics, we focus on the physics of blockade effect on the quantum many-body ground states by preparing ordered phases of the 1D chain and 2D square lattice. The associated Jupyter notebook is provided here.

**Best practices tip:**
When choosing the number of measurements per task, there is a trade-off between shot noise and speed/cost. We find that ~100 measurements are a good middle ground, with up to 1000 shots needed for high-resolution phase diagrams and low-probability outcomes, and as few as 25 shots for parallelized arrays.

As we understand from previous examples, an intuitive understanding of the blockade effect on the ground state is as follows: With a positive detuning Δ, adding Rydberg excitations lowers the ground state energy. However, the Rydberg excitations are subject to the Rydberg blockade effect—a large positive energy penalty which enforces that only one Rydberg excitation is allowed within the blockade radius. The interplay of these two mechanisms allows the creation of different ordered states (and more exotic states) depending on the strength of the blockade radius and the detuning. Let us start with the 1D chain for simplicity: if the blockade radius encompasses nearest neighbors, the many-body ground state is an alternating pattern of ground and Rydberg state, called a $Z_2$ state. If the blockade radius contains next-nearest neighbors, the many-body ground state is alternating ground-ground-Rydberg, or $Z_3$; and so on.

## 4.1. The 1D $Z_2$ phase

We use an adiabatic evolution to prepare the simplest quantum many-body ground state, the 1D $Z_2$ state. To do that, we can start with all atoms in the ground state |0⟩, which is the ground state of the many-body Hamiltonian with a large negative detuning Δ. Then, the Rabi frequency Ω is turned on, and the detuning strength is ramped up from a large negative value to positive values. If such a process is slow enough, the quantum state of the system stays close to the ground state of the time-dependent Hamiltonian at time $t$ by the adiabatic theorem. At the end of this process, we arrive at a target Hamiltonian, and correspondingly, the prepared state is the ground state of the final Hamiltonian. One example of an adiabatic protocol is shown in Fig. 4.1.

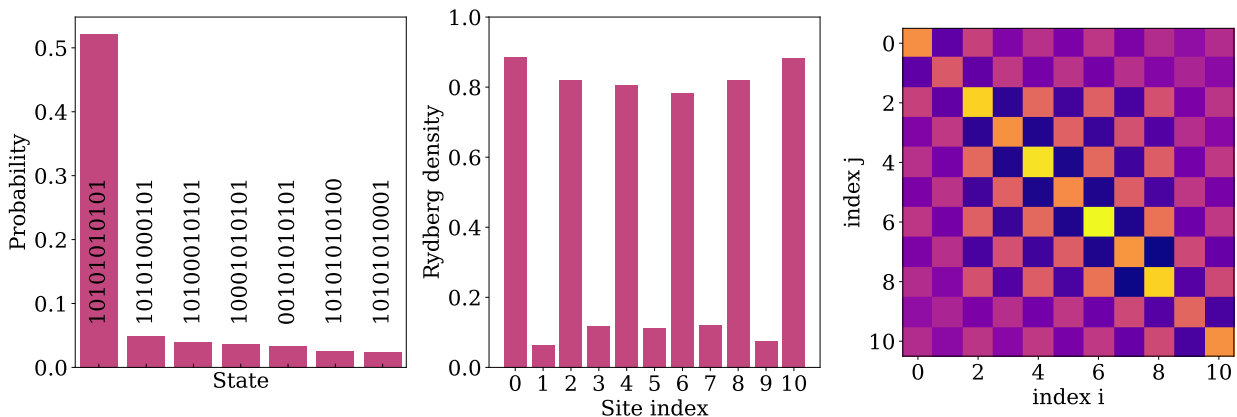

Figure 4.1 The bitstring distribution after state preparation for a 4 μs total evolution time for a 11-site 1D chain. Of the 377 measurements, we find 200 (53%) are the ground state; the next most probable bitstrings all have a single excitation missing, consistent with an ≈ 8% error in detecting the Rydberg state. Middle: Rydberg density per site, indicating the clear staggered signature of a $Z_2$ state. Right: the two-point connected correlation function $C_{ij}$; we find a correlation length of ≈ 3.6 sites, which is on par with [Bernien2017].

The left panel of Figure 4.1 shows the resulting probability for the bitstring distribution for 11 atoms and a total time of 4 $\mu s$. The final state has a large proportion of target $Z_2$ state, where there is one Rydberg excitation in every other atom. We find an $\approx 58\%$ $Z_2$-state probability, which is slightly larger than analogous sizes compared with [Bernien2017]. Most of the infidelity can be accounted for by considering that there is a finite measurement error of $\approx 8\%$ of spuriously measuring a Rydberg state as a ground state; in this case, the probability of reading out 10101010101 given a measurement of that state is $(1 - 0.08)^6 \approx 60\%$. One can also clearly see the pattern of the $Z_2$ state by looking at the average value of Rydberg excitation at each site, as shown in the middle plot in Figure 4.1, which alternates between occupied (1) and unoccupied (0).

> **Best practices tip:**
> When deciding a direction for a 1D chain, it is best for it to be vertical (Y direction). The Rydberg lasers propagate in this direction, and so vertical chains are less sensitive to any spatial inhomogeneities.

For a perfect $Z_2$ state, any two atoms will be correlated even if they are infinitely far away. However, experimentally prepared states will always have other configurations, resulting in a finite correlation length. To quantitatively characterize the quality of the prepared state, we use the connected two-point correlation function,

$$C_{ij} = \langle n_i n_j \rangle - \langle n_i \rangle \langle n_j \rangle,$$

where the average is taken over experimental repetitions. Figure 4.1 right shows the correlation function between each two pair of sites, which demonstrates clear antiferromagnetic order. The correlation length $\lambda$ can be extracted by fitting $C_{ij} \sim \exp(-|i - j|/\lambda)$ and find $\lambda \approx 3.6$ sites. This value is on par with state-of-the-art performance [Bernien2017]. Note that even though the correlation length is relatively small, the fidelity can still be large, because bulk defects must always come in pairs. Defects come in indistinguishable pairs, so the probability of no defects is $\left(1 - \frac{1}{2} 0.25^2\right)^N \approx 0.97^N$, which is a smaller penalty than measurement error $\approx 0.92^{N/2}$.

## 4.2. Adiabatic preparation performance characterization

Finally, we can study fidelity of the 1D $Z_2$ states as a function of total evolution time and system size. Figure 4.2 left plots the state fidelity as a function of total evolution time for a fixed size $N = 11$. In the noiseless theory limit, the probability increases monotonically as a function of the preparation time, consistent with the adiabatic limit. However, due to decoherence effects, the fidelity will decay at large times. Compellingly, we find that fidelity does not decay significantly even for times of order 10 $\mu s$, suggesting that evolution is fully coherent within the 4 $\mu s$ window accessible at Amazon Braket. The right half of Figure 4.2 shows the state fidelity as a function of total chain length, from $N = 5$ to $N = 19$ for a fixed evolution time $T = 7\mu s$. As expected, the fidelity drops exponentially, due to the finite correlation length and measurement error. We find a $Z_2$ state probability of $\approx 40\%$ for the chain of length $N = 19$. The purple line is the numerical expectation scaled by an 8% measurement error $0.92^{(N+1)/2}$.

> **Best practices tip:**
> There is a relatively large ($\sim 8\%$) error of incorrectly measuring a Rydberg state as a ground state. When comparing theory to experimental data, it is important to include this dominant effect, as shown in Fig. 4.2 by the purple line.

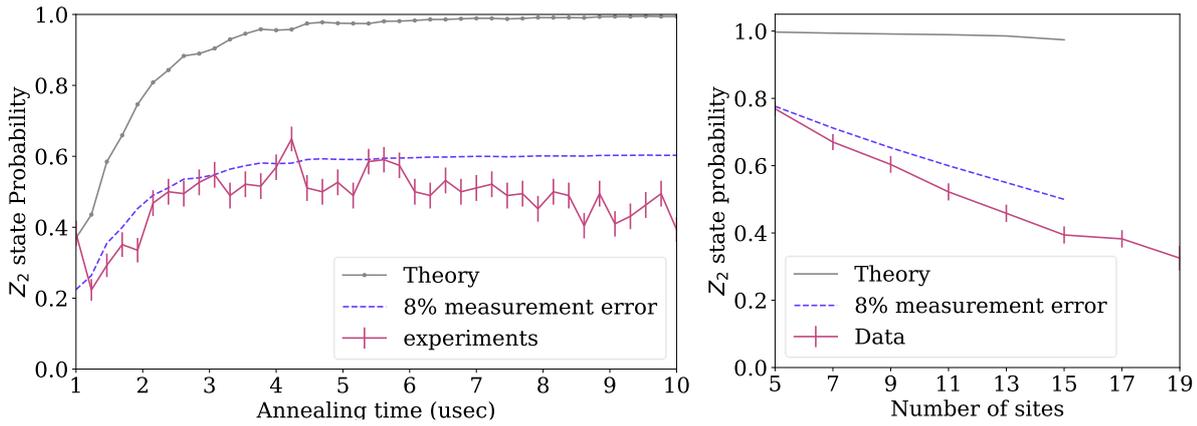

Figure 4.2 Left: Ground state fidelity as a function of total annealing time for fixed $N = 11$. Red is the probability of experimentally measuring the $\cdots 10101 \cdots$ bitstring. Black is the noiseless theory expectation, while purple is the numerical expectation scaled by an 8% measurement error of spuriously measuring a Rydberg state as ground. For short annealing times, finite-speed diabatic errors dominate, reducing the probability. for longer annealing times, the noiseless system asymptotically approaches unit fidelity but the experimental system decoheres, though persists in quality even for long times. Right: Ground state fidelity as a function of system size for a fixed time $T = 7 \, \mu s$. We find a fidelity of $F \approx 40\%$ for an $N = 19$ chain, which is comparable with the state-of-the-art performance. Note that the maximum time and Y extent ($6.1 \, \mu m \ast 19 = 115.9 \, \mu m > 76 \, \mu m$) of these protocols may be larger than the available range on Amazon Braket and is available in premium access mode.

## 4.3. The 2D striated and checkerboard phase

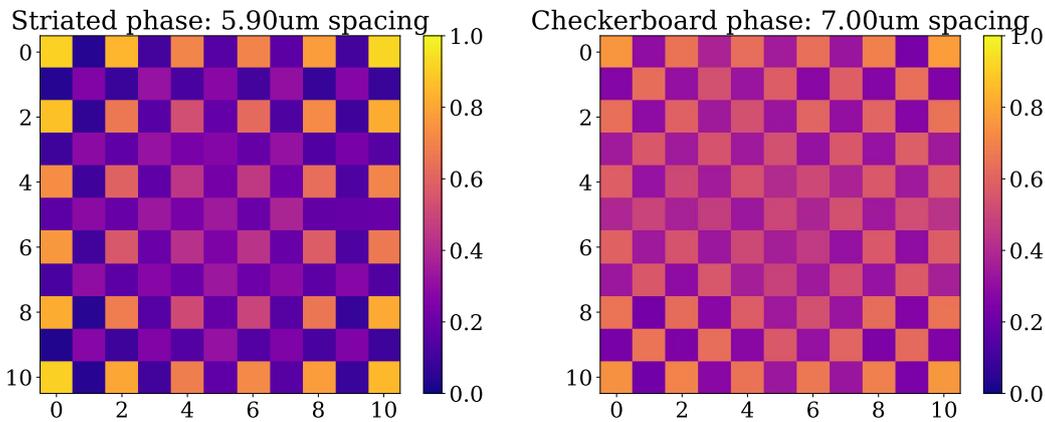

Figure 4.3 Rydberg density on each site for the striated phase (left) and checkerboard phase (right). Each pixel is an atom on the grid, with the color representing the average Rydberg density. The states are pinned to a single configuration by boundary effects due to the odd extent. Observe that the striated phase has an even superposition ($\sim 0.5$ density) of states in the bulk, which is an indicator of a quantum phase. Similarly, observe that the checkerboard phase has alternating occupied-unoccupied sites similar to the 1D $Z_2$ phase.

Aquila offers flexibility of controlling the geometry of atoms in 2D as well. Below, we will show the results of using Aquila to prepare ground states of 2D square lattices. The ground state physics of 2D Rydberg atom arrays has been theoretically studied [Samajdar2020] and later experimentally probed [Ebadi2020]. In this example, we prepare the checkerboard state and the striated phase (a phase with strong quantum fluctuations) for a 2D $11 \times 11$ grid of atoms. Like 1D $Z_2$ state preparation case, we also use the quantum adiabatic evolution to prepare these striated phase and checkerboard phase, where the detuning starts from negative and adiabatically ramps to the positive target value. As shown in [Ebadi2020], the ratio between blockade radius and the lattice constant is different for the two phases. We will change the lattice constant to make sure the parameters fit into the two different phase regimes.

Figure 4.3 plots averaged Rydberg density for the two prepared states, which differ only by the lattice spacing of $a = 5.9 \ \mu m$ and $7.0 \ \mu m$. As we see, the checkerboard phase features alternating Rydberg excitations, like the 1D case we have seen above. However, the striated phase features a strong uncertainty for atoms that sit in the middle of four Rydberg excitations. So, the average Rydberg density for these sites is intermediate between 0 and 1, which is an indication that this is a state with no classical counterpart. Indeed, one can show that the phase is induced by quantum fluctuation given by the Rabi (transverse) term. These are all strong demonstrations of 2D many-body quantum phases. For more details, we encourage the reader to read [Samajdar2020].

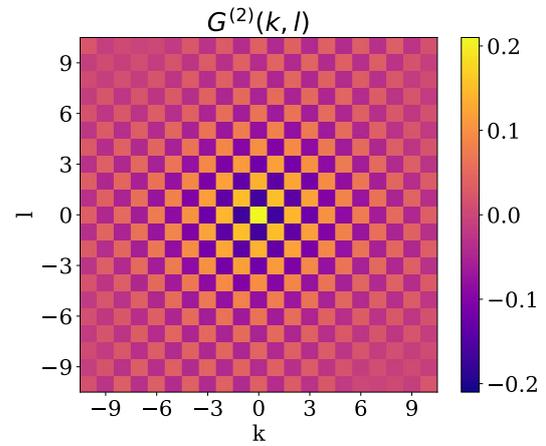

Figure 4.4. Two-site correlation function for the checkerboard phase of Fig. 4.3 Right, showing clear antiferromagnetic order. The correlation function decays exponentially with distance from the origin with a correlation length of $\lambda \approx 5.7$.

Finally, Figure 4.4 plots the connected two-point correlation function for the checkerboard state, which is computed as

$$G^{(2)}(k, l) = \frac{1}{N_{(k,l)}} \sum_{ij} \left( \langle n_i n_j \rangle - \langle n_i \rangle \langle n_j \rangle \right),$$

where the sum is over all pairs of atoms $(i, j)$ separated by the same distance $(k, l)$ sites. In the perfect limit, there is no decay and $G^{(2)}(k, l) = -1^{k+l}$, but due to noise and decoherence the correlation function falls exponentially. Like the 1D $Z_2$ case, we see a long-range correlation between pairs of sites. We find the correlation length to be $\approx 5.7$ sites.

**Best practices tip:**
When preparing problems with many atoms, it is important to post-select on the atom arrays being fully loaded, otherwise one might find unexpected data. Atoms are loaded into sites with a probability > 99%.

# 5. Example 4: Many-body quantum scars

In the previous example, we have seen how to use adiabatic evolution to prepare the $Z_2$ state in a 1D atom chain. A previous experimental study [Bernien2017] discovered that if one implements non-equilibrium dynamics with a $Z_2$ state of a 1D chain, the Rydberg blockade constraint results into persistent revivals of quantum dynamics, in contrast to the expectation of reaching thermalization quickly. Later theoretical studies [Turner2018] revealed that this behavior is due to special eigenstates embedded in the quantum many-body spectrum, a phenomenon called quantum many-body scars. **Quantum many-body scars** are analogous to the phenomenon of classical scars in single-particle quantum chaos, where scars represent a concentration of some eigenfunctions along the trajectory of classical periodic orbits. Similarly, in the quantum many-body case, the initial $Z_2$ state has a large overlap with these specific scar states. Under the time evolution of the Rydberg Hamiltonian, the initial state undergoes the trajectory of periodic quantum orbits. The non-thermal behavior is thus caused by **non-ergodicity** in the Hilbert space.

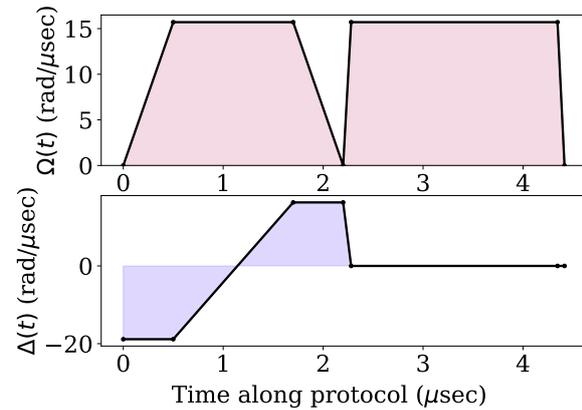

Figure 5.1 Protocol for many-body scars. The first half of evolution adiabatically prepares a $Z_2$ state, while the second half is a non-equilibrium quench under the scar Hamiltonian.

In this example, we use Aquila to simulate quantum many-body scar dynamics. We demonstrate the persistent revivals of many-body dynamics with measurements of the Rydberg density. The program happens in two parts. In the first part, the initial state is prepared using an adiabatic protocol, and in the second part the state undergoes a quench and implements non-equilibrium dynamics. An example Jupyter notebook is provided here.

Figure 5.1 shows the full waveform for the dynamics for the adiabatic state preparation. It consists of two parts — the first part is the quantum adiabatic evolution for preparation of the $Z_2$ state (as we have seen from the previous example), and the second part is the waveform for evolution of quantum scar, where the detuning has been quenched to zero. Results for the protocol are shown in Fig. 5.2. The top right panel in Figure 5.2 plots the average Rydberg density as a function of time. The black curve shows the classical simulation results, and the red curves show the results from Aquila, with a good agreement between the two. Dynamics of average density are clearly shown reflecting the two parts of the protocol. In the first part, the average density increases to a finite value due to the adiabatic preparation of the $Z_2$ state. In the second phase, the density shows coherent evolution oscillating between two extrema with a characteristic frequency for long times, instead of quickly reaching a steady value (thermalization). This is a strong indication of the quantum scar dynamics.

To further visualize the dynamics, we track the probability for one of the Neel states, 10101010101, as a function of time, as shown in the top left panel of Fig. 5.2. In the preparation part, the probability increases to a large value. In the second part after the quench, the states show clear revival behavior: after certain fixed time, the quantum state has a large overlap with the tracked Neel state. This is consistent with quantum many-body scar picture where the quantum states go through a periodic trajectory in the Hilbert space. Finally, the bottom panel in Fig. 5.2 plots the site-resolved real-time evolution of the Rydberg density, where the vertical axis indexes the atom position and horizontal indexes evolution time. Once again, it confirms the expectation for the quantum scar dynamics that the state oscillates between two Neel states 10101010101 and 01010101010. For more details about quantum many-body scar in constrained model, we encourage the readers to read [Turner2018].

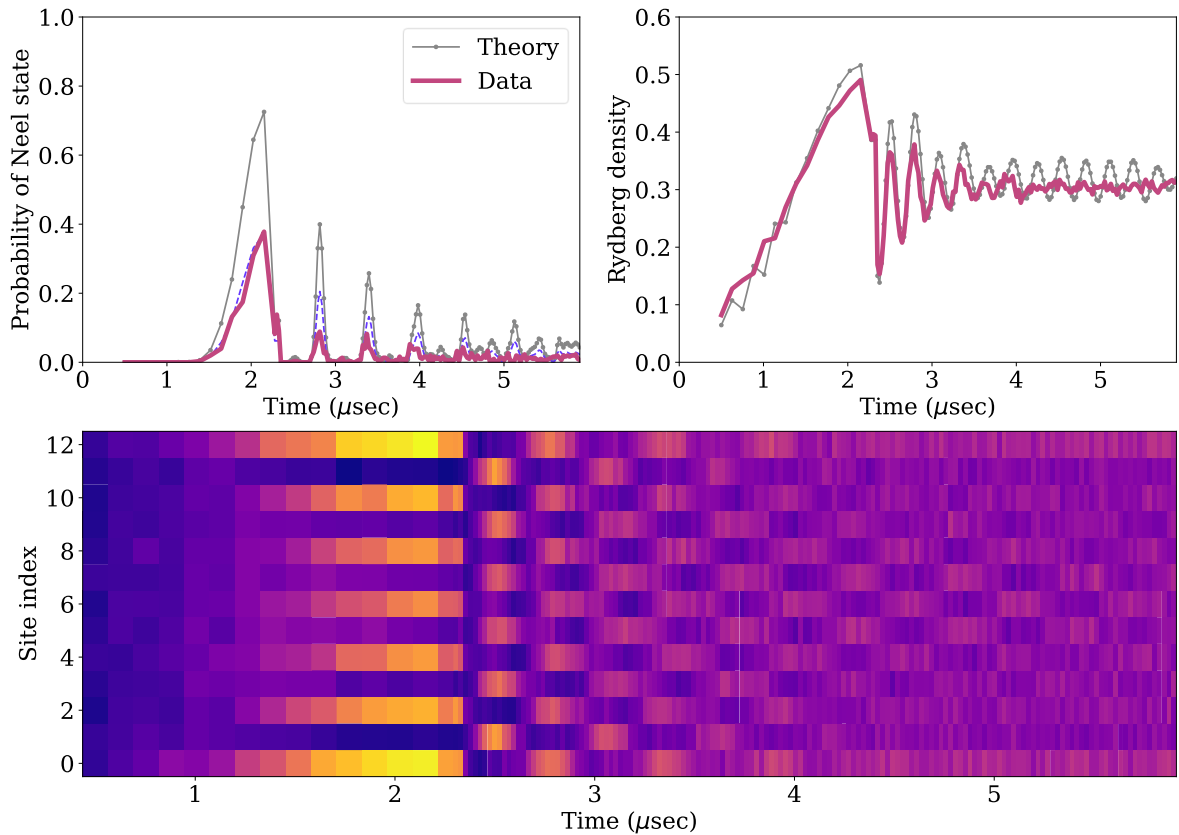

Figure 5.2. 1D quantum many-body scar dynamics on 13 sites with adiabatic state preparation. Top left plots the probability of the Neel state 1010101010101 as a function of evolution time. The spikes after the quench are a clear indicator of revivals and athermal behavior; dashed purple line is numerical simulations scaled by the 92% Rydberg state measurement fidelity. Top right plots the Rydberg density as a function of evolution time. The long-time oscillations are another indicator of athermal behavior. Bottom plots the Rydberg density for all sites. After the quench, the state oscillates between the 10101 and 01010 scar states, as expected. Note that the maximum time is greater than the 4 $\mu s$ restriction on Amazon Braket and is available in premium access mode.

# 6. Example 5: Maximum independent set on unit disk graphs

The final example minimally reproduces the recent seminal work [Ebadi2022], which included researchers at QuEra and solves the **maximum independent set** (MIS) problem on unit disk graphs. The accompanying Jupyter notebook is hosted here. The MIS problem is as follows: given some graph of vertices $V$ and edges $E$, find the largest subset of vertices $I \subset V$ such that no two vertices of $I$ are connected by an edge. This is a prototypical hard optimization problem, and in different contexts is NP-Hard, NP-complete, and average-case hard. A subset of graphs is called unit disk graphs, where each vertex has a position in 2D space, and there is an edge between a pair of vertices if and only if they are closer than some radius $R_{ud}$, called the **unit disk radius**. This is still a hard optimization problem and is NP-complete under a quadratic reduction to MIS on general graphs.

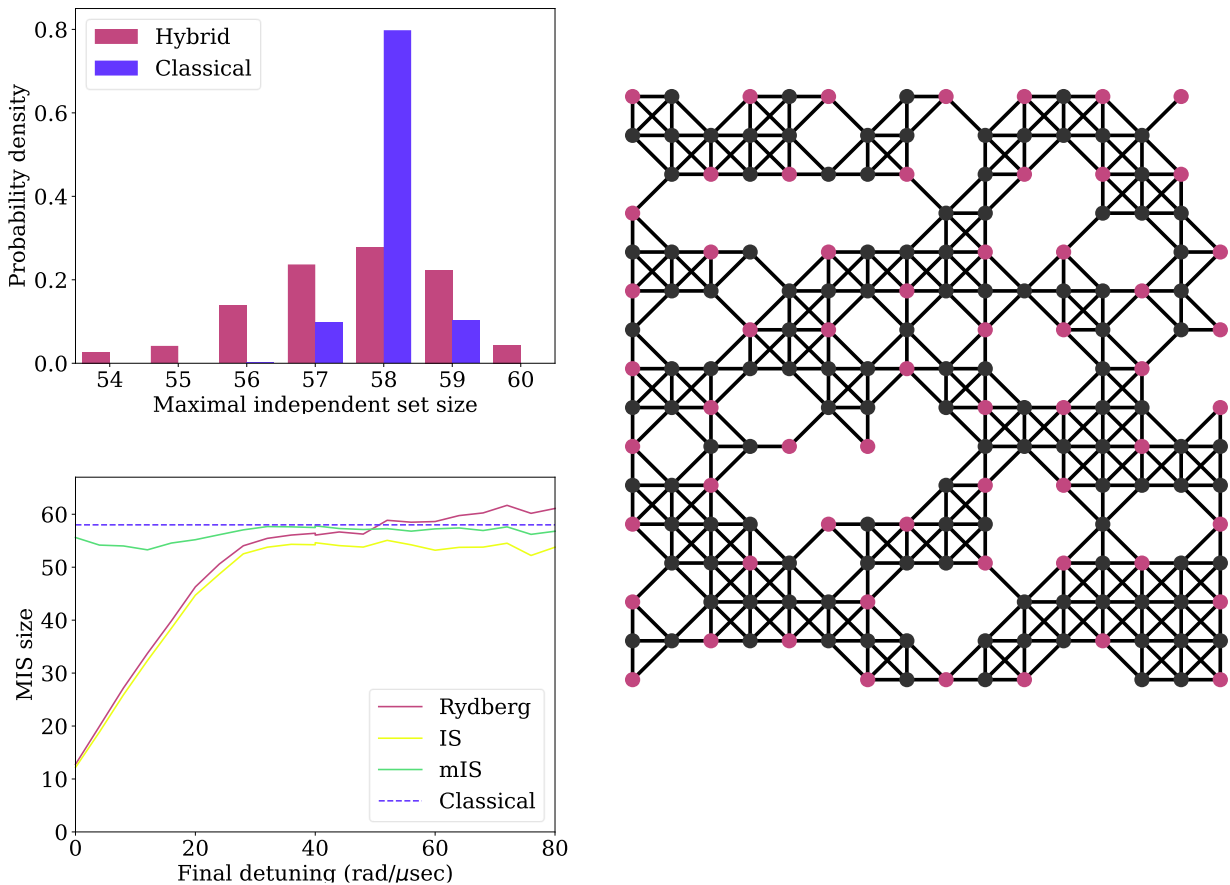

Figure 6.1. Variational optimization and performance of adiabatic protocols. The protocol is a simple linear adiabatic ramp where the final detuning is chosen to be the single variational parameter. The performance as a function of final detuning is plotted on the bottom left. Red plots the average number of Rydberg excitations on the graph; yellow is after minimally removing independent set violations; green is after greedily adding vertices until the set is maximal. Purple is the performance of the classical-only algorithm, which serves as a baseline. Observe that below the detuning corresponding to the unit disk radius, of $\sim 30 \frac{rad}{\mu s}$, there is a large drop-off in the independent set size. top right shows the target King's graph with a 30% dropout, with an MIS of size 60. left plots the distribution of mIS for the hybrid vs. classical algorithm for a final detuning of $\Delta_f = 40 \frac{rad}{\mu s}$. Observe that the greedy classical algorithm does very well, finding an average mIS of 58.0. The hybrid algorithm finds an average mIS of 57.5 but has a higher chance of finding a MIS than the classical algorithm. This data is from 200 shots.

Computing the MIS on these unit disk graphs can be solved on neutral-atom hardware by linking the unit disk radius to the blockade radius $R_{ud} = R_b$, so that the ground state of the Hamiltonian encodes the MIS by measuring which atoms are excited to the Rydberg state. The graph is directly encoded onto hardware by placing an atom at every (scaled) position of the logical target graph, and the state is prepared using the quantum adiabatic algorithm. The protocol used is a simple linear ramp like that shown in Fig. 3.1. The Rabi frequency is ramped to a value of $\Omega = 15 \frac{rad}{\mu s}$, and the detuning is ramped from an initial value of $\Delta = -30 \frac{rad}{\mu s}$ to some final value $\Delta_f$. The final value is a free variational parameter, which is directly optimized on the machine on a single graph instance using a 1D grid search with $\Delta_f \in [\,0, 80\,] \frac{rad}{\mu s}$. The logical target graph is a King's graph with a 30% random dropout, where each vertex is adjacent with the same connectivity as that of the King's moves on a chessboard. An example $16 \times 16$ graph with 183 vertices is shown in Fig. 6.1. The lattice spacing between adjacent vertices is 5 $\mu m$, and so the unit disk radius is $R_{ud} \in (5\sqrt{2}, 10)\ \mu m$. The detuning linked to the blockade radius of the unit disk radius is between $5.4 < \Delta_b < 43 \frac{rad}{\mu s}$. For more details, see [Ebadi2022].

While the annealing schedule on Aquila is coherent through the full 4 $\mu s$ evolution, it is not perfect due to diabatic, noise and decoherence effects. This may result in measurements which violate the independent set condition or may not be maximal independent sets (mIS). A mIS is an independent set that cannot be added with additional vertices without violating the independent set constraint, but it may not be a maximum independent set, which is the largest independent set. To improve the measurements, a minimal classical post-processing procedure is used. First, a greedy algorithm minimally removes independent set violations by finding a greedy IS on the subgraph induced by the Rydberg measurements. Then, the same greedy algorithm adds vertices to the independent set, if able, to find a mIS. For more details on these algorithms, see the example Jupyter notebooks here.

The behavior of this post-processing procedure can be seen in Fig. 6.1, where the average set size over post-processed measurements on a single graph is plotted as a function of final detuning. For small $\Delta_f$, the Rydberg count (red) is much less than the MIS of 60 corresponding to a large effective blockade radius; for large $\Delta_f$, the count is on average larger, corresponding to a small effective blockade radius and thus many independent set violations. The greedy-removed average IS size (yellow) is always smaller than both the MIS and the Rydberg density and has a turnover from the Rydberg size around the detuning corresponding to the blockade radius. Finally, the greedy-add average mIS (green) is larger than the IS (yellow) but may be smaller than the Rydberg count. To distinguish quantum performance from classical-only performance, the algorithm can also be run on the classically sampleable "all zeros" or "all ones" bitstring, which finds an average mIS of 58.0; note that this is equivalent to quantum performance for a protocol with $\Delta_f \ll 0$ or $\Delta_f \gg 0$ respectively. By optimizing $\Delta_f$ over the post-processed objective function of average mIS size, we find an optimal value of $\Delta_f \approx 40 \frac{rad}{\mu s}$.

Performance of the hybrid algorithm is shown in Fig. 6.1 Bottom left, which plots the distribution of maximal independent set sizes. Observe that in Fig. 6.1 bottom the classical algorithm does quite well on this particular graph, often finding an mIS of size 58 or 59. The average mIS from the classical algorithm is 58.0. This is better on average than the hybrid algorithm, which has an average mIS of 57.5. However, there is a small ($\sim 4.5\%$ probability) of the hybrid algorithm to find the MIS of size 60, which is heuristically larger than that of the classical algorithm. This data is from 200 shots on the machine, plus 10 rounds of post-processing per shot.

While the performance is quite promising on this graph, it is important to confirm that this behavior extends to an ensemble of graphs. One characterization of the relative performance between algorithms is given by the performance ratio $PR$, which in this case is the average mIS found by the hybrid algorithm, divided by the average mIS found by the classical-only algorithm [Wurtz2021]. If $PR = 1$, then the two algorithms have equivalent performance, if $PR > 1$, then the hybrid algorithm has better performance than classical. If the average $\langle PR \rangle > 1$, there is an ensemble advantage of the hybrid algorithm in comparison to the classical algorithm. Data over an ensemble of 50 graphs is shown in Fig. 6.2; curiously, we find that the average is less than classical only, with only a few atypical problem instances having any performance boost. This finding is consistent with [Ebadi2022], which characterizes the hardness by the degeneracy ratio of the number of first excited states over the number of MIS.

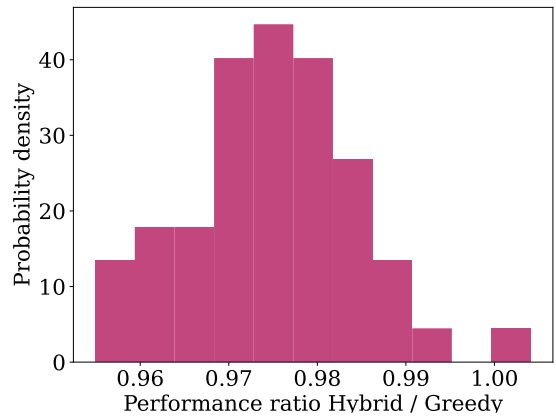

Figure 6.2. Distribution of performance ratio of quantum vs. classical average mIS size over an ensemble of 50 graphs. We find an average $\langle PR \rangle = 0.974$, indicating no ensemble advantage.

## About QuEra

Located in Boston, QuEra Computing is a maker of advanced quantum computers based on neutral atoms, pushing the boundaries of what is possible in the industry. Founded in 2018, the company is built on pioneering research recently conducted nearby at both Harvard University and MIT. QuEra is building the industry's most scalable quantum computers to tackle useful but classically intractable problems for commercially relevant applications. Our signature machine, Aquila, is available now for general use over the Amazon Braket cloud.

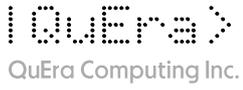
QuEra Computing Inc.

QuEra Computing Inc.
1284 Soldiers Field Road
Boston, MA 02135

T +1 617-588-7207
info@quera.com
quera.com